\documentclass[acmsmall]{acmart}

\AtBeginDocument{%
  \providecommand\BibTeX{{%
    \normalfont B\kern-0.5em{\scshape i\kern-0.25em b}\kern-0.8em\TeX}}}

\setcopyright{acmlicensed}
\copyrightyear{2023}
\acmYear{2023}
\acmDOI{XXXXXXX.XXXXXXX}

\acmJournal{JACM}
\acmVolume{XX}
\acmNumber{X}
\acmArticle{1}
\acmMonth{12}

\usepackage{caption}
\usepackage{algpseudocode}
\usepackage{amsmath}
\usepackage{amsfonts}
\usepackage{subfigure}
\usepackage{graphicx}

\usepackage{amsfonts}
\usepackage{booktabs}
\usepackage{siunitx}

\usepackage{color}

\usepackage{tikz}

\usepackage{enumitem}

\usepackage{url}
\makeatletter
\def\url@leostyle{%
  \@ifundefined{selectfont}{\def\UrlFont{\sf}}{\def\UrlFont{\small\ttfamily}}}
\makeatother
\newcommand{\eat}[1]{}

\usepackage[linesnumbered,ruled]{algorithm2e}

\usepackage{mdframed} 

\usepackage{xcolor}
\definecolor{light-gray}{gray}{0.9}

\newenvironment{packed_enum}{%
  \begin{enumerate}%
  }{\end{enumerate}}

\usepackage{amsthm}

\newcolumntype{L}[1]{>{\raggedright\let\newline\\\arraybackslash\hspace{0pt}}m{#1}}

\begin{document}


\title{Blockchain Takeovers in Web 3.0: \\
An Empirical Study on the TRON-Steem Incident 
}


\author{Chao Li}
\affiliation{%
  \institution{Beijing Key Laboratory of Security and Privacy in Intelligent Transportation, Beijing Jiaotong University}
  \city{Beijing}
  \country{China}
}
\email{li.chao@bjtu.edu.cn}

\author{Runhua Xu}
\affiliation{%
  \institution{School of Computer Science and Engineering, Beihang University}
  \city{Beijing}
  \country{China}
}
\email{runhua@buaa.edu.cn}

\author{Balaji Palanisamy}
\affiliation{%
  \institution{School of Computing and Information, University of Pittsburgh}
  \city{Pittsburgh}
  \country{USA}
}
\email{bpalan@pitt.edu}

\author{Li Duan}
\affiliation{%
  \institution{Beijing Key Laboratory of Security and Privacy in Intelligent Transportation, Beijing Jiaotong University}
  \city{Beijing}
  \country{China}
}
\email{duanli@bjtu.edu.cn}

\author{Meng Shen}
\affiliation{%
  \institution{School of Cyberspace Security, Beijing Institute of Technology}
  \city{Beijing}
  \country{China}
}
\email{shenmeng@bit.edu.cn}

\author{Jiqiang Liu}
\affiliation{%
  \institution{Beijing Key Laboratory of Security and Privacy in Intelligent Transportation, Beijing Jiaotong University}
  \city{Beijing}
  \country{China}
}
\email{jqliu@bjtu.edu.cn}

\author{Wei Wang}
\affiliation{%
  \institution{Beijing Key Laboratory of Security and Privacy in Intelligent Transportation, Beijing Jiaotong University}
  \city{Beijing}
  \country{China}
}
\email{wangwei1@bjtu.edu.cn}

\renewcommand{\shortauthors}{Chao et al.}

\begin{abstract}

A fundamental goal of Web 3.0 is to establish a decentralized network and application ecosystem, thereby enabling users to retain control over their data while promoting value exchange. However, the recent Tron-Steem takeover incident poses a significant threat to this vision. In this paper, we present a thorough empirical analysis of the Tron-Steem takeover incident. By conducting a fine-grained reconstruction of the stake and election snapshots within the Steem blockchain, one of the most prominent social-oriented blockchains, we quantify the marked shifts in decentralization pre and post the takeover incident, highlighting the severe threat that blockchain network takeovers pose to the decentralization principle of Web 3.0. Moreover, by employing heuristic methods to identify anomalous voters and conducting clustering analyses on voter behaviors, we unveil the underlying mechanics of takeover strategies employed in the Tron-Steem incident and suggest potential mitigation strategies, which contribute to the enhanced resistance of Web 3.0 networks against similar threats in the future. We believe the insights gleaned from this research help illuminate the challenges imposed by blockchain network takeovers in the Web 3.0 era, suggest ways to foster the development of decentralized technologies and governance, as well as to enhance the protection of Web 3.0 user rights.

\end{abstract}




\maketitle

\section{Introduction}

Web 3.0, also known as the decentralized Web, is widely recognized as the next stage in the evolution of the Web~\cite{alabdulwahhab2018web,shen2024artificial}.
It aspires to remodel our digital space into a user-centric environment, where individuals retain comprehensive ownership and control over their personal data~\cite{chen2022digital,yang2023zero}.
This next-generation Web, distinct from the static content delivery of Web 1.0 and the interactive and collaborative characteristics of Web 2.0, is marked by its commitment to decentralized applications and reinforced security.
\textcolor{black}{Web 3.0 is powered by the integration of multiple cutting-edge information and communication technologies, including blockchain~\cite{guidi2020steem,li2019incentivized,su2023hybrid,shi2023ress} and social computing~\cite{guidi2020graph,li2023liquid,guidi2020blockchain}. }
These innovations together facilitate a network and application ecosystem underpinned by decentralization~\cite{lin2021measuring,li2023cross}, where the exchange of value is not only possible but also streamlined, and the balance of power between users and digital service providers could be realigned~\cite{wu2023financial,guidi2022assessment}.

However, as with any significant paradigm shift, the evolution towards Web 3.0 presents a set of new challenges and unprecedented hurdles~\cite{ragnedda2019blockchain,chen2022digital}.
Among these emergent concerns, one of the most pressing is the threat of hostile takeovers, a term usually associated with the world of business acquisitions~\cite{shivdasani1993board,franks1996hostile}.
Historically, a hostile takeover refers to a scenario where an individual or organization acquires a target company against the will of the target company's management.
In the context of Web 3.0, the notion of hostile takeovers evolves to reflect the distinctive characteristics of this novel digital ecosystem. 
Instead of traditional centralized companies in the physical world, the potential targets of hostile takeovers in Web 3.0 have converted into decentralized blockchain networks~\cite{leiponen2022dapp,ba2022fork,li2023hard}.

As a fundamental building block of Web 3.0, blockchain is a distributed ledger technology that leverages peer-to-peer (P2P) networks, fueled by cryptographic algorithms
and consensus protocols (e.g., PoW~\cite{nakamoto2008bitcoin}, PoS~\cite{king2012ppcoin}, DPoS~\cite{li2023characterizing}), to record and verify transactions in a decentralized manner. 
The decentralized nature of blockchain networks is the key to their resilience against censorship and single-point-of-failure attacks, offering a viable alternative to the centralized systems that dominate the Web 2.0 era~\cite{leiponen2022dapp}.
However, the decentralized nature of blockchain networks also makes them vulnerable to hostile takeovers, as the lack of a central authority makes it difficult to prevent a hostile entity from gaining control of the network.
In the context of blockchain, a hostile takeover refers to a scenario where an entity acquires the governance or decision-making power of a target blockchain network against the general will of the block producers in charge of managing the target blockchain network~\cite{li2023hard}.
Such a takeover would result in a significant change in the strategic direction of the target blockchain network, which could pose a severe threat to the network's integrity and the interests of its users, and even lead to network fragmentation, also known as a hard fork~\cite{ba2022fork}.

Delegated Proof-of-Stake (DPoS)~\cite{larimer2014delegated} is one of the most widely adopted consensus protocols in the Web 3.0 era, powering many popular blockchain networks, including EOSIO~\cite{liu2022decentralization}, Tron~\cite{tron}, and Steem~\cite{li2019incentivized}. 
DPoS has also given rise to a series of improved variants, such as Liquid Proof-of-Stake (LPoS)~\cite{allombert2019introduction}, Nominated Proof-of-Stake (NPoS)~\cite{wood2016polkadot}, and Proof of Staked Authority (PoSA)~\cite{wang2022exploring}, which greatly enrich the diversity of the Web 3.0 ecosystem.
Compared to Proof-of-Work (PoW) blockchains, which often suffer from low transaction throughput~\cite{croman2016scaling}, DPoS strikes a unique balance between decentralization and scalability by introducing a voting-based governance model~\cite{li2023characterizing}. 
In this model, the governance power is distributed among a set of block producers (BPs) elected by coin holders via stake-weighted votes, enabling a more efficient and scalable blockchain network that satisfies the needs of Web 3.0 applications~\cite{li2023hard}.
However, the coin-based voting governance model also makes DPoS blockchains vulnerable to hostile takeovers, as the governance power of a DPoS blockchain is concentrated in the hands of a small number of BPs~\cite{li2023hard,liu2022decentralization,guidi2022assessment,guidi2021analysis}. 
In fact, there have been several real-world attempts where DPoS blockchains were targeted for takeover by hostile entities. 
The most notable one is the TRON-Steem takeover incident in 2020, where the TRON Founder acquired Steemit Inc., the company behind one of the most prominent social-oriented blockchain named the Steem blockchain, and used its stake in Steem to replace the original BPs with its own BPs~\cite{li2023hard,ba2022fork}. 
This compelled numerous core users to exit the Steem blockchain network they had cultivated for years, severely undermining the ecosystem of Steem, a quintessential Web 3.0 network.
As a result, the hostile takeover triggered a hard fork of the Steem blockchain, giving birth to a new blockchain network called Hive~\cite{Hive}. 
This incident has raised serious concerns about the security of DPoS blockchains and the Web 3.0 ecosystem as a whole~\cite{li2023hard,ba2022fork}.

In this paper, we present a large-scale longitudinal study of the TRON-Steem takeover incident.
The dataset used in this study includes 41,818,752 Steem blocks prior to the Hive Fork, as well as 10,675,297 Steem blocks and 10,811,841 Hive blocks after the Fork. 
The primary objectives of this study are twofold: to empirically demonstrate the impact of blockchain network takeovers on decentralization, and to unveil the underlying mechanics of takeover strategies.

Decentralization is a fundamental principle of Web 3.0, which enables users to have greater control over their digital assets and promotes transparency and accountability in the network~\cite{chen2022digital,wu2023financial}. 
This principle is critical to the development of a more sustainable and collaborative Web 3.0 ecosystem, where users can participate in the governance and decision-making processes of the network~\cite{kiayias2022sok}. 
In this study, we aim to investigate the impact of the TRON-Steem takeover incident on the decentralization of the Steem blockchain. 
\textcolor{black}{To achieve this, we estimate the previously unrecorded VESTS/STEEM exchange rate, a critical parameter for calculating the voting power derived from the coins staked by voters, and reconstruct the historical stake and election snapshots of the Steem blockchain over a period of five years.}
We propose a novel measurement algorithm called Voter-layer Decentralization Quantification (VLDQ) to quantify the decentralization of DPoS blockchains at the voter layer. 
We then apply VLDQ to the Steem and Hive blockchains to quantify the decentralization of these two blockchain networks before and after the Hive Fork.
Our results show that the hostile takeover by TRON has significantly decreased the decentralization of the Steem blockchain, which is inconsistent with the general perception of the Steem community. 
\textcolor{black}{Furthermore, we find that Hive tends to be more decentralized than Steem was after the takeover, yet it does not reach the level of decentralization that Steem maintained prior to the takeover, which may suggest a potential long-term negative impact on decentralization. }
These findings highlight the importance of protecting the decentralization of blockchain networks from hostile takeovers.

In addition to investigating the impact of hostile takeovers on decentralization, we also delve into the underlying mechanics of takeover strategies.
We propose heuristic methods to identify anomalous voters and conduct clustering analyses on voter behaviors.
Specifically, we first extract distinct voting strategies performed by voters and obtain the various combinations of block producers (BPs) that any voter has chosen in the history of BP election.
We then track the history of switching voting strategies for each selected voter, which reveals the voter's voting strategy switching pattern in terms of both the switching time and the switching direction.
Based on the switching patterns, we employ heuristic methods to identify anomalous voters who suddenly begin to participate in the BP election after a long absence. 
We then conduct clustering analyses on the switching patterns of all selected voters to identify clusters of voters who have switched voting strategies in the same way, as well as their strategy switching patterns and proxying relationships.
By delving deeper into the takeover strategies employed during the incident, we suggest potential mitigation strategies that can be applied to fortify future Web 3.0 networks against such threats.

\noindent \textbf{Contributions. }
In a nutshell, this paper makes the following key
contributions:

\begin{itemize}[leftmargin=*] 
\item \textcolor{black}{ We present a large-scale longitudinal study of the TRON-Steem takeover incident, employing a rich dataset that includes a vast number of blocks from both pre-fork and post-fork periods. }
\item \textcolor{black}{ We develope the VLDQ algorithm to measure DPoS blockchain decentralization at the voter layer and show that the TRON-Steem takeover significantly reduced it, highlighting the risks of hostile takeovers to Web 3.0.}
\item \textcolor{black}{ We propose heuristic methods to identify voter behavior patterns and develop mitigation strategies, enhancing Web 3.0 network security against hostile takeovers and providing insights for future research.}
\end{itemize}

The rest of this paper is organized as follows: 
We start by introducing the background in Section~\ref{sec2} and data collection in Section~\ref{sec3}.
In Section~\ref{sec4}, we focus on measuring decentralization. We first fix the missing system parameter, then reconstruct daily stake snapshots and finally presents the results and findings at the BP layer and the voter layer, respectively.
In Section~\ref{sec5}, we focus on analyzing the takeover strategies. We first extract distinct voting strategies, then identify anomalous voters and conduct clustering analyses on voter behaviors.
We discuss related work in Section~\ref{sec6} and conclude in Section~\ref{sec7}.

\section{Background}
\label{sec2}

In this section, we first provides an overview of the Steem blockchain~\cite{guidi2020blockchain,li2019incentivized,li2021steemops}, including its most popular Web 3.0 application \textit{Steemit}, its implementation of the DPoS consensus protocol and its ecosystem in general.
We then provide a more detailed account of the TRON-Steem takeover incident~\cite{li2023hard,ba2022fork}, including its timeline and key events.

\subsection{Steem Blockchain Network}

Steem, similar to Ethereum~\cite{wood2014ethereum} and EOSIO~\cite{liu2022decentralization}, is a blockchain supporting a wide range of Web 3.0 applications.
To date, there have been over 324 Steem-based decentralized applications\footnote[1]{https://steem.com/developers/}, with most of these applications designed to serve social users in the Web 3.0 era.
Among them, the first and most popular application is \textit{Steemit}, which functions as a Web 3.0 version of Reddit.
In \textit{Steemit}, users have the ability to generate and circulate content in the form of blog posts, which can then receive responses, reposts, upvotes, or downvotes.
Importantly, each vote carries a weight corresponding to the stake of the voter. Periodically, coins are apportioned as rewards to posts receiving the highest ranks, which incentivizes users to generate high-quality content , functioning as a decentralized content curation system for Web 3.0.
The Steem blockchain keeps a record of the data generated by Steem-based Web 3.0 applications and creates a new block every three seconds to log all operations verified by block producers (BPs).

With a user base of over a million, Steem has recorded nearly a billion operations, demonstrating the scalability achievable on a Web 3.0 platform.
Through the implementation of the Delegated Proof-of-Stake (DPoS) consensus protocol~\cite{larimer2014delegated}, Steem illustrates an effective means of distributing authority and power within a network for large-scale Web 3.0 applications, where traditional Proof-of-Work (PoW) blockchains may encounter scalability issues.
Under the DPoS system, coin holders can cast up to 30 stake-weighted votes for block producer (BP) candidates, with the top-20 becoming BPs responsible for managing high-quality servers to produce blocks in turn.
BPs also make decisions on the governance of the network, such as updating system parameters, adopting new features, and even blocking suspicious accounts.
In Steem, any proposal related to governance must be approved by a supermajority of BPs, defined as 17 out of 20 BPs.
Alternatively, users may set proxies to allow other users to make all BP election votes on their behalf. The proxy feature adds a layer of complexity to the reconstruction of historical election snapshots due to the potential for creating a chain of proxies.
This delegation and election process demonstrates the decentralized governance ideal for Web 3.0, where users have direct influence over the network's operation.

Next, for the sake of simplicity, we present a simplified version of Steem's coin ecosystem.
Similar to most blockchains, Steem issues native coins known as \textit{STEEM}. 
To cast stake-weighted votes, a coin holder can stake, aka lock, \textit{STEEM} to receive voting power, denoted as \textit{VESTS}, at a rate $\lambda$ \textit{VESTS}/\textit{STEEM}.
However, the raw blockchain data does not directly provide the rate $\lambda$.
Coin holder can withdraw their invested \textit{STEEM} whenever they wish, but the invested fund is automatically partitioned into thirteen segments, to be withdrawn over a thirteen-week period.
Notably, Steem allows for the purchase of \textit{VESTS} by one user (A) for another user (B), in which case A loses \textit{STEEM} while B gains \textit{VESTS}. 
Likewise, Steem permits a user (A) to withdraw invested \textit{STEEM} to another user (B), in which case A loses \textit{VESTS} while B gains \textit{STEEM}. 
These dynamic vesting methods, combined with the missing system parameter $\lambda$, further increase the difficulty of reconstructing historical stake snapshots.

\begin{figure}
  \centering
  {
      \includegraphics[width=0.7\columnwidth]{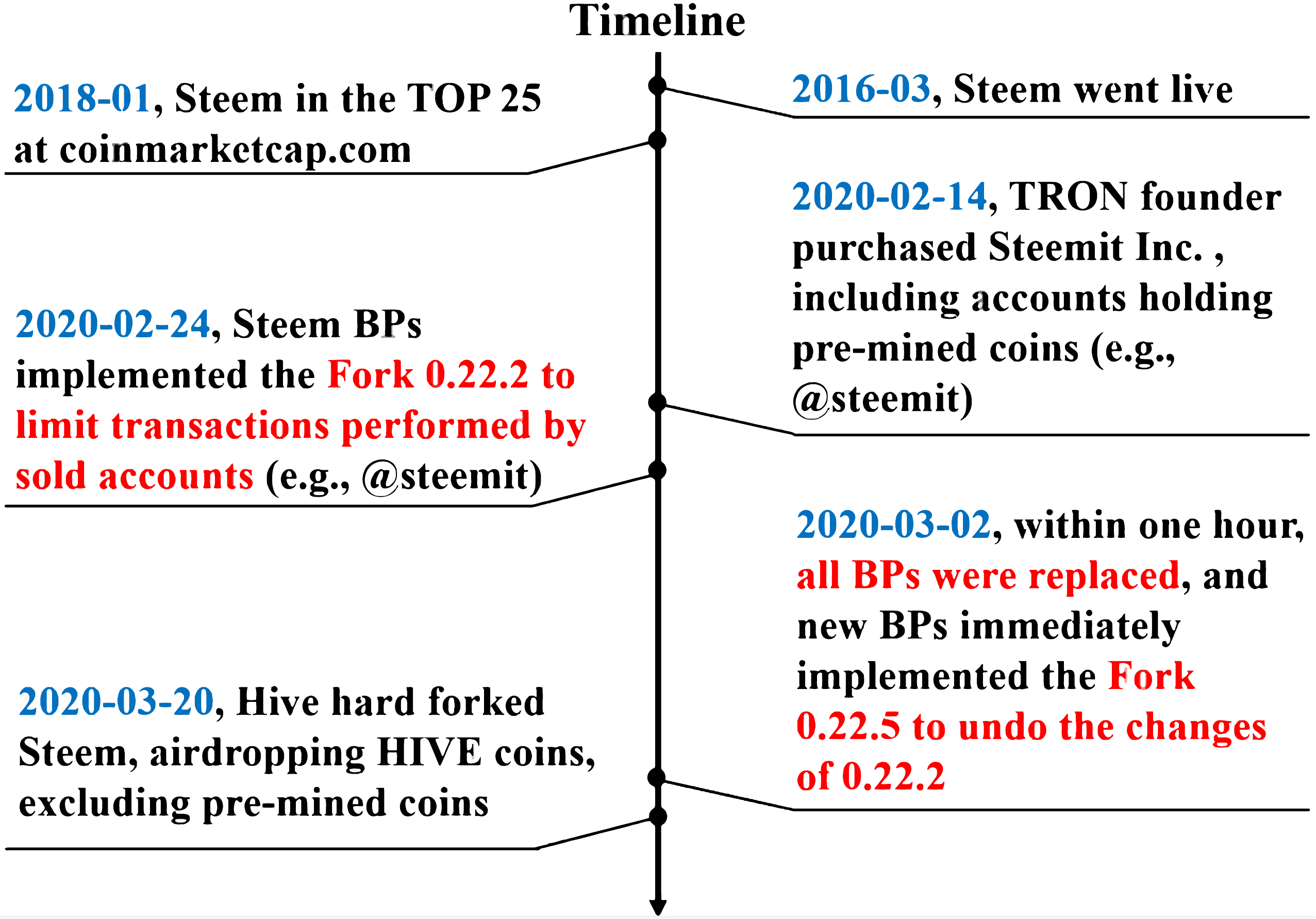}
  }
  \caption {Timeline for TRON-Steem takeover incident}
  \label{front_page} 
\end{figure}

\subsection{TRON-Steem Takeover Incident}


\textcolor{black}{The TRON's takeover of Steem, spearheaded by the founder of TRON, was primarily motivated by the desire to integrate Steem’s successful Web 3.0 social ecosystem into TRON's expanding network. This move aimed to leverage Steem's established user base and technological advancements to bolster TRON's position in the decentralized social media space, reflecting a strategic push for market expansion.}

In early 2020, the founder of TRON purchased Steemit Inc. and gained control of their pre-mined coins, which initiated the TRON-Steem takeover incident as summarized in Fig.~\ref{front_page}. 
Prior to the purchase, Steemit Inc. had promised that the pre-mined coins, which accounted for approximately 20\% of the STEEM supply, would be non-voting stake. 
However, the details of how these coins would be used after the purchase had not been agreed upon. 
As a result, top BPs implemented Fork 0.22.2\footnote[2]{https://steemit.com/steem/@softfork222/soft-fork-222} to enshrine the promises in code, which restricted the abilities of five relevant accounts, including \textit{@steemit}, in terms of transferring coins and participating in BP election. 
However, TRON founder quickly took action against Fork 0.22.2. 
On March 2nd, 2020, within one hour, all of the top-20 BPs were suddenly replaced by newcomers, who then immediately revoked the restrictions via Fork 0.22.5~\footnote[3]{https://github.com/steemit/steem/pull/3618}. 
Ultimately, the original BPs decided to move to Hive, a blockchain that hard forked Steem and excluded the pre-mined coins, rather than continue to fight a losing battle. 

The TRON-Steem takeover incident represents the first de facto takeover of a decentralized blockchain by centralized capital in the Web 3.0 era. 
However, little is known about the impact of the hostile takeover on decentralization, as well as the underlying mechanics of takeover strategies.
Next, after introducing the data collection in Section~\ref{sec3}, we answer the two questions in Section~\ref{sec4} and~\ref{sec5}, respectively.

\section{Data collection}
\label{sec3}
In this section, we describe our data collection methodology. 
The Steem blockchain provides an Interactive Application Programming Interface (API) for developers and researchers to collect and parse blockchain data~\cite{SteemAPI}. 
Similarly, the Hive blockchain offers an API at~\cite{HiveAPI}. 
Since Hive hard forked Steem at block 41,818,753, we first collected blocks 1 to 41,818,752, which were shared by both Steem and Hive before the fork. 
These blocks were produced between the inception of Steem (2016-03-24) and the Hive fork day (2020-03-20). 
To investigate the post-Hive-fork era, we collected blocks created between 2020-03-20 and 2021-03-31, namely block 41,818,753 to 52,494,049 from Steem and block 41,818,753 to 52,630,593 from Hive, respectively. 
Interestingly, we found that Hive produced 136,544 more blocks than Steem during the same period of time after the hard fork. 
On average, there should be 28,800 new blocks generated per day, so the difference may suggest that Steem BPs missed about 360 more blocks than Hive BPs per day after the hard fork.

\section{Understanding the Decentralization}
\label{sec4}


In this section, our objective is to investigate and empirically demonstrate the impact of the hostile takeover on decentralization. 
\textcolor{black}{To achieve this, we first fix the missing system parameter, the VESTS/STEEM exchange rate $\lambda$, and reconstruct the daily stake snapshots, which are essential for measuring decentralization.
Then, based on a layered measurement model, we compare the decentralization of Steem and Hive before and after the takeover at both the BP layer and the voter layer. }
Our study on decentralization aims to provide a comprehensive understanding of the context of decentralization and the changes in decentralization resulting from the takeover.



\subsection{Reconstructing Daily Stake Snapshots}
\label{stake_snapshot}

\noindent \textbf{Methodology }
In Steem, the stake, which represents the amount of voting power (or \textit{VESTS}) held by a voter at a certain block height $h$, can be computed as follows:
\begin{equation}
v_h = \sum_{i=1}^{h} (\lambda \cdot v^{t2v}_{i}+v^{r}_{i}-v^{fvw}_{i}) \label{e1}
\end{equation}
Here, $\sum_{i=1}^{h} \lambda \cdot v^{t2v}_{i}$ represents the total amount of \textit{VESTS} purchased using the coin \textit{STEEM} (recorded in operations such as \textit{transfer\_to\_vesting} in blocks), 
$\sum_{i=1}^{h} v^{r}_{i}$ represents the total amount of \textit{VESTS} rewarded by the platform (recorded in four types of operations, e.g., \textit{curation\_reward} and \textit{author\_reward}), 
and $\sum_{i=1}^{h} v^{fvw}_{i}$ represents the total amount of \textit{VESTS} withdrawn from the platform (recorded in operations such as \textit{fill\_vesting\_withdraw}).

\begin{figure}
\centering
{
    \includegraphics[width=0.6\columnwidth]{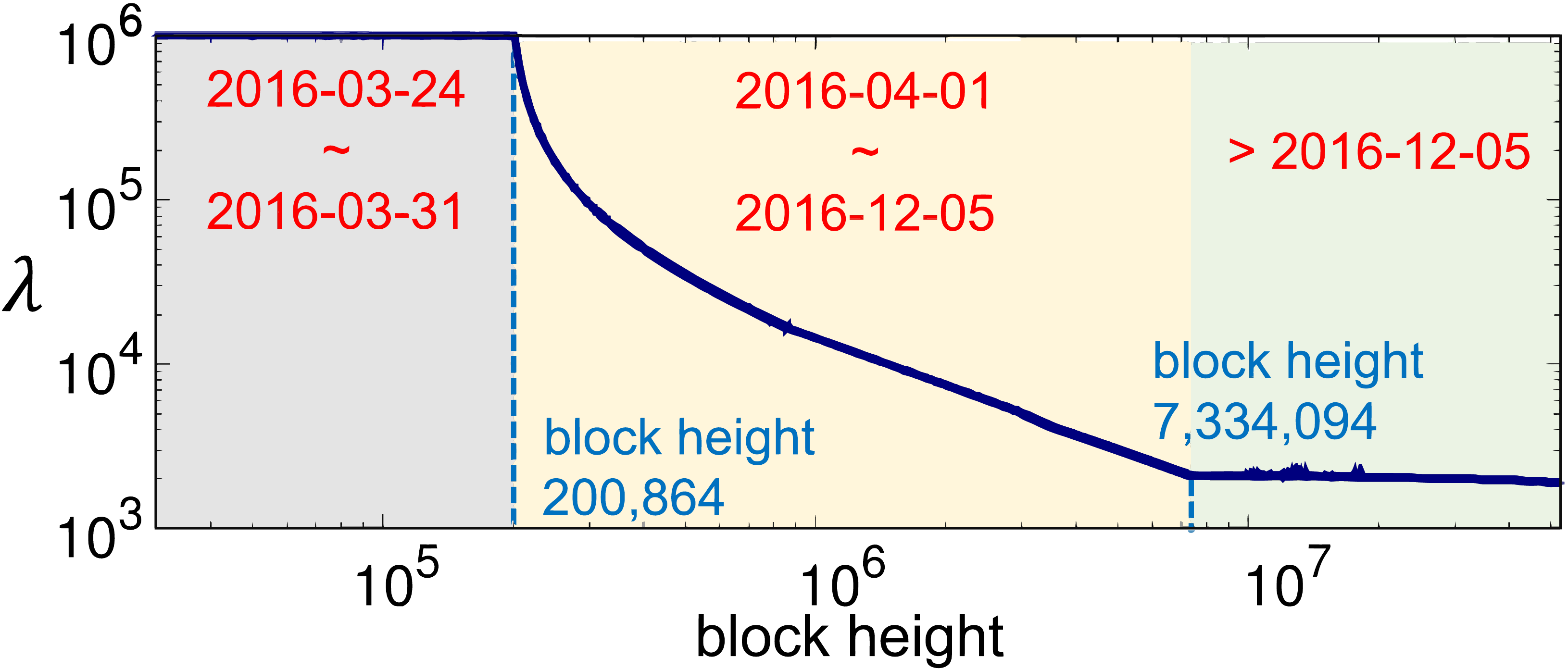}
}
\caption {\small Values estimated for VESTS/STEEM exchange rate $\lambda$ from block 1 to block 52,630,593}
\label{lambda} 
\end{figure}

It is not difficult to extract historical $v^{t2v}_{i}$, $v^{r}_{i}$ and $v^{fvw}_{i}$ from blockchain raw data.
However, the \textit{VESTS}/\textit{STEEM} exchange rate $\lambda$ is missing in raw data collected from~\cite{SteemAPI}, as an operation \textit{transfer\_to\_vesting} only discloses the amount of invested \textit{STEEM}, but reveals no information about the amount of gained \textit{VESTS}.
To compute $\lambda$, we investigated two helpful Steemit blogs \footnote[4]{https://steemit.com/steemdev/@jesta/historical-rates-for-vests-and-steem.}\footnote[5]{https://steemit.com/steem/@crokkon/historic-rates-for-steem-per-vests-2018-2019ytd.}
that discuss the missing parameter $\lambda$. 
Both blogs indicated that parameter $\lambda$ could be computed through operations \textit{fill\_vesting\_withdraw}, which are used by voters to withdraw \textit{VESTS}.
An operation \textit{fill\_vesting\_withdraw} provides both the amount of withdrawn \textit{VESTS} and the corresponding amount of deposited \textit{STEEM}, so the ratio of the two values gives $\lambda$.
However, the first \textit{fill\_vesting\_withdraw} operation was performed at block height 479,660, while the first \textit{transfer\_to\_vesting} operation was performed at block height 28,361.
Unfortunately, a value for $\lambda$ estimated at a certain block height could only reflect the exchange rate at that moment.
As a result, we do not know the values for $\lambda$ before block 479,600, namely the period between 2016-03-24 and 2016-04-10, and we could not accurately estimate historical voting power for any voter who has performed operation \textit{transfer\_to\_vesting} before day 2016-04-10.

To overcome this challenge, we explored alternative approaches.
Fortunately, we discovered that the Hive blockchain has recently introduced a new type of operation, \textit{transfer\_to\_vesting\_completed}~\footnote[6]{https://gitlab.syncad.com/hive/hive/-/issues/111.}
in 2021, which provides additional information about the amount of gained \textit{VESTS} in conjunction with the \textit{transfer\_to\_vesting} operation. This information enables us to estimate $\lambda$ before day 2016-04-10. The results of our fine-grained estimation for $\lambda$ are presented in Fig.~\ref{lambda}.

\begin{figure}
  \centering
  {
      \includegraphics[width=0.6\columnwidth]{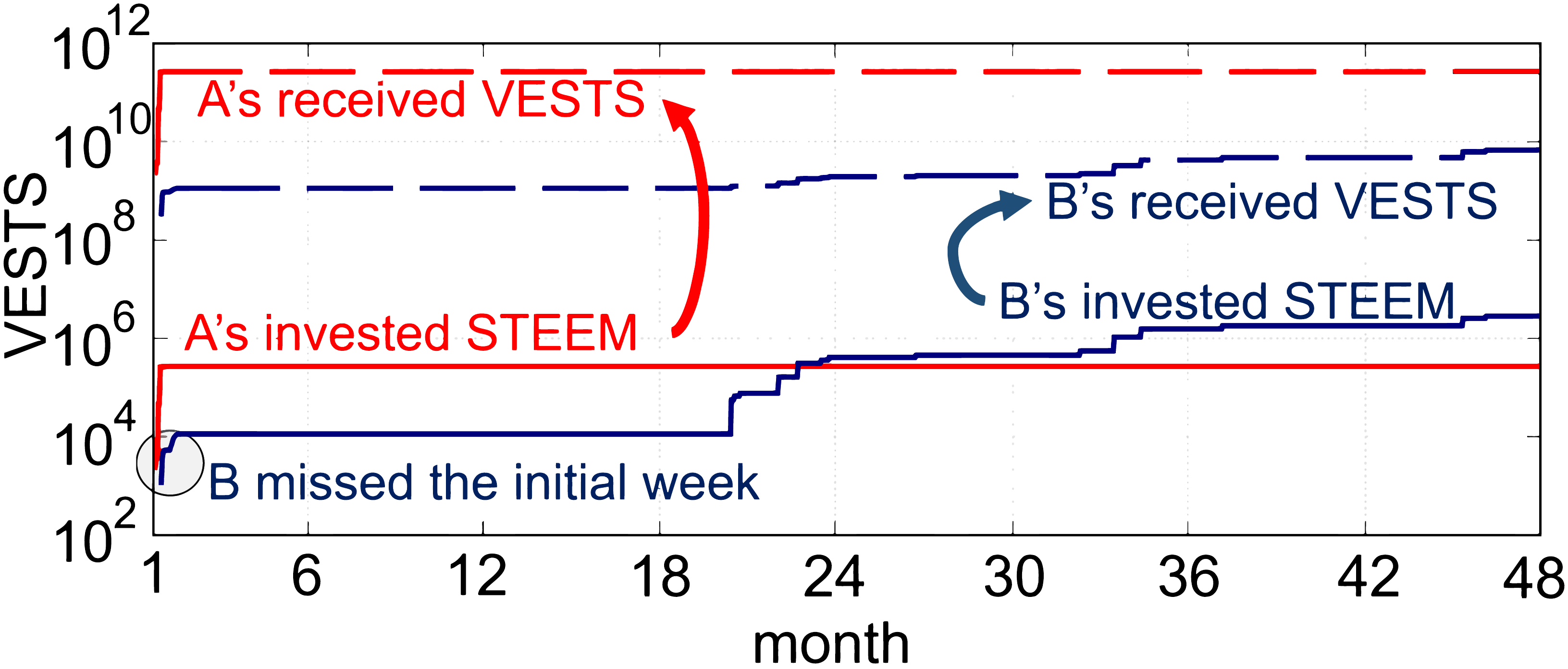}
  }
  \caption {\small A real example illustrating the significant impact of $\lambda$. 
  Voter A purchased VESTS mainly during the first week (before 2016/03/31), while voter B purchased VESTS after 2016/03/31
  }
  \label{sec5_1_02} 
  \end{figure}

\noindent \textbf{Observation }
Fig.~\ref{lambda} illustrates the significant changes in the value of the exchange rate $\lambda$ over time. 
Surprisingly, during the first week of Steem's operation, from 2016-03-24 to 2016-03-31, we observe a huge value of $\lambda$, namely 1,000,000. 
However, from 2016-12-06 to the present, each \textit{STEEM} can only purchase around 2,000 \textit{VESTS}.
Between 2016-04-01 and 2016-12-05, the value of $\lambda$ dropped by 99.8\%.
These findings suggest that the exchange rate $\lambda$ has undergone significant decrease over time.

To illustrate the significant impact of $\lambda$ on reconstructing historical stake snapshots, we present a real example in Fig.~\ref{sec5_1_02} between two voters, A and B. 
In this example, voter A purchased \textit{VESTS} mainly during the first week (before 2016-03-31), while voter B purchased \textit{VESTS} many times, but only after 2016-03-31. 
We can observe that, even though voter B gradually invested over 10 times the amount of \textit{STEEM} invested by A, in month 48, A held about 39 times the amount of \textit{VESTS} possessed by B.

\begin{figure}
  \centering
  {
      \includegraphics[width=0.7\columnwidth]{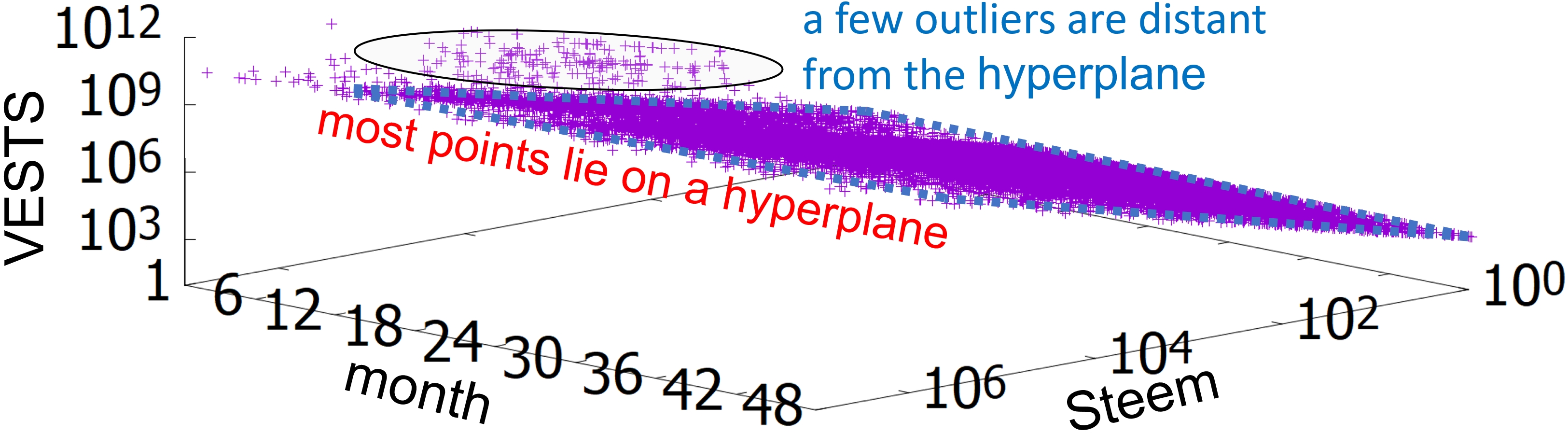}
  }
  \caption {\small A scatter plot characterizing voters using three variables, the total amount of invested \textit{STEEM}, the total amount of purchased \textit{VESTS} and the first date of investment}
  \label{sec5_1_03} 
\end{figure}

To investigate the impact of early investment on the voting power of voters, we present a scatter plot in Fig.~\ref{sec5_1_03}, where each point represents a voter who has cast at least one vote to BP candidates. 
Each point presents three variables corresponding to a voter, including the total amount of invested \textit{STEEM}, the total amount of purchased \textit{VESTS}, and the date of the first investment. 
The results show that most points are located on a hyperplane, which suggests that most voters did not purchase their \textit{VESTS} in the early days and have a ratio of their total \textit{VESTS} over total \textit{STEEM} close to 2,000. 
However, we do observe a group of outliers located far from the hyperplane. 
Nearly all these outliers performed their first \textit{transfer\_to\_vesting} operation during the first week and have a very high \textit{VESTS}/\textit{STEEM} ratio.

\noindent \textbf{Insights }
The significant fluctuations in the VESTS/STEEM exchange rate $\lambda$ suggest that early investors may have benefited more from their investments.
Interestingly, this phenomenon is similar to the business investment model, where angel investors provide capital for start-ups.
However, in the context of decentralization, the results suggest that it becomes increasingly difficult for latecomers to compete with early investors in Steem.
Furthermore, the results also indicate the challenge for later block producers to compete with the stake that TRON founder acquired from Steemit Inc., the founder and earliest investor of Steem.

\subsection{Layered Measurement Model}
\label{metrics}

Recent research on decentralization measurement in blockchains has revealed that decentralization measured at different layers can vary significantly~\cite{zeng2021characterizing,li2020comparison,li2023cross}.
This is due to the different distribution of resources across different layers, such as the distribution of computational power among mining pools and pool participants in PoW blockchains~\cite{zeng2021characterizing}, as well as the distribution of voting power among BPs and voters in DPoS blockchains~\cite{li2020comparison}.
To address this issue, we propose a layered measurement model that measures decentralization at two layers: the BP layer and the voter layer.

Intuitively, decentralization at the BP layer measures the distribution of voting power received by BPs from voters, while decentralization at the voter layer measures the distribution of voting power among voters.
However, in practice, BPs in Steem and Hive create blocks in turn regardless of the voting power they receive from voters once they enter the top-20.
Additionally, voters with the same amount of voting power may have different impacts on the block-creation competition due to the different number of votes they cast and whether their selected BPs are in the top-20.
\textcolor{black}{Therefore, in the next two sections, we measure decentralization at the BP layer and the voter layer, respectively.}

\subsection{BP-layer Decentralization}
\label{pool}

\begin{figure}
  \centering
  {
      \includegraphics[width=0.60\columnwidth]{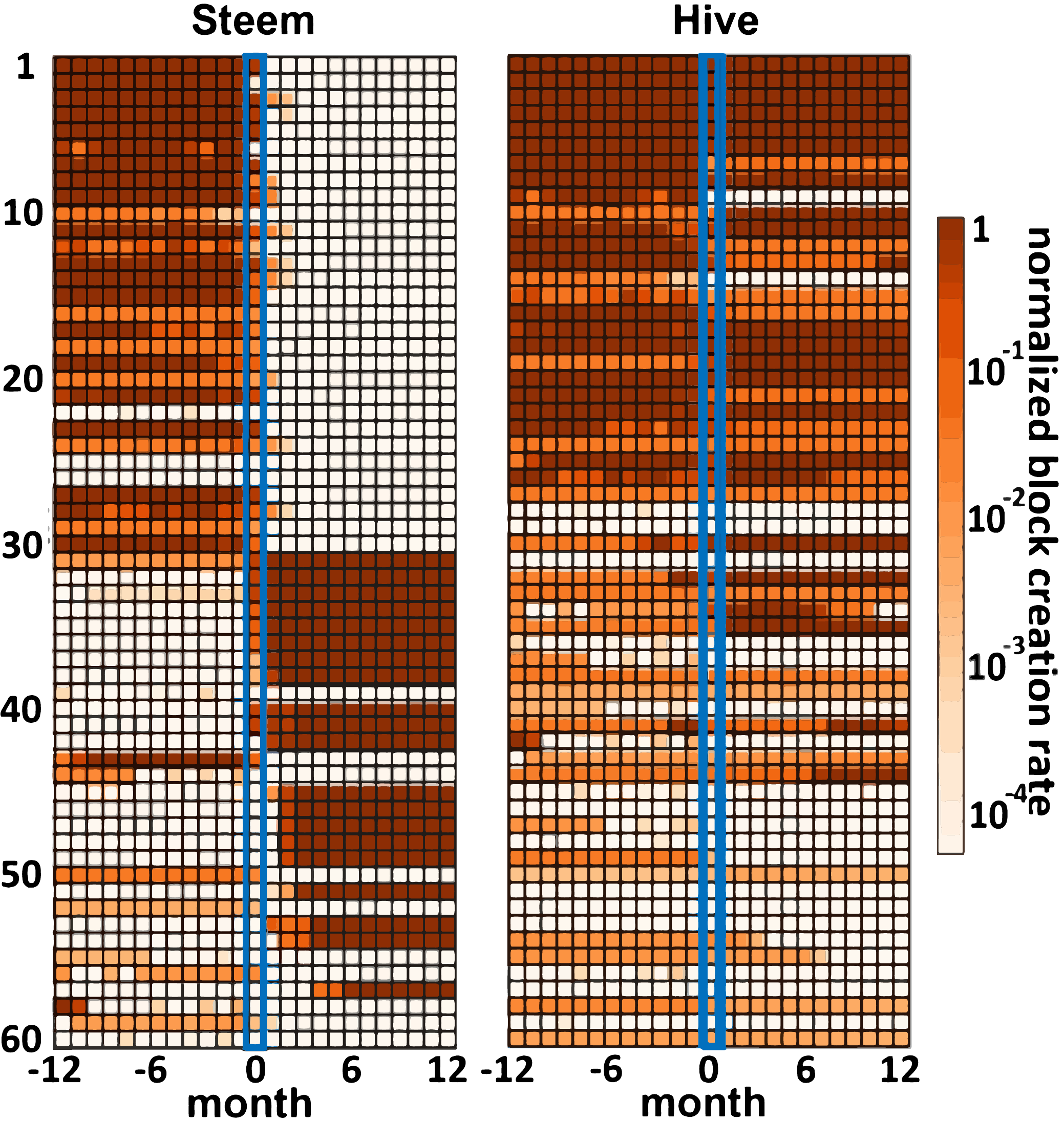}
  }
  \caption {Heatmap of top-60 BPs' normalized block creation rates in Steem and Hive during the takeover month (month 0), one year before the takeover (month -12 to -1) and one year after the takeover (month 1 to 12)}
  \label{heatmap_2} 
\end{figure}

In this section, \textcolor{black}{to investigate BP-layer decentralization, we have developed Algorithm~\ref{A1} to measure the block production rates of the most influential block producers (BPs) by parsing the raw blockchain data and ranking the BPs based on the number of blocks they have created.
Then, to effectively highlight variations and trends in BP activity over time and offer insights into the decentralization dynamics, we visualize the outputs of this algorithm as a heatmap in Fig.~\ref{heatmap_2}.}

\begin{algorithm}
  \footnotesize
  \SetKwInOut{Input}{Input}
  \SetKwInOut{Output}{Output}

  \Input{Blockchain raw data $\mathbf{D}$, start/end block height $\{s, e\}$, BP range $l$, month range $m$.}
  \Output{Normalized block creation rate $Q = \{q_{i}^{j}|1 \leqslant i \leqslant l, 1 \leqslant j \leqslant m\}$.}

  $B = \{b_h|s \leqslant h \leqslant e\} \gets parseBP(\mathbf{D})$\;
  $N = \{n_i^j | 1 \leqslant i \leqslant l, 1 \leqslant j \leqslant m, 
    \sum_{j=1}^{m}(n_{i+1}^j) \leqslant  \sum_{j=1}^{m}(n_i^j)\} 
    \gets rankBP(B)$\;
  \For{$i = 1\ to\ l, j = 1\ to\ m$} 
  {
    $q_{i}^{j} = \frac{n_{i}^{j}}{Max(N)}$\;
  }
  
  \caption{BP-layer Decentralization Measurement}
  \label{A1} 
\end{algorithm}


\noindent \textbf{Methodology }
In Algorithm~\ref{A1}, we initiate the process by invoking the $parseBP()$ function to identify the block producer (BP), $b_h$, of each block from the raw block data of the blockchain, denoted as $\mathbf{D}$ (line 1).
We then use the $rankBP()$ function to select the top-$l$ BPs that have generated the most blocks during a selected number of $m$ months in three steps (line 2): 
(1) quantify the number of blocks created by each distinct BP per month; 
(2) rank all BPs based on the total amount of blocks created by each of them during $m$ selected months; 
(3) select the top-$l$ BPs that have created the largest amount of blocks during the $m$ selected months.
The output of $rankBP()$ includes the number of blocks created by each of the top-$l$ BPs during each of the $m$ months, denoted as $N$, providing a comprehensive view of the block production activities of the most influential BPs.
Finally, we compute the per-month normalized block creation rate $q_i^j$ for each of the top-$l$ BPs during the $m$ months by dividing the number of blocks created by the $i^{th}$ BP during the $j^{th}$ month, denoted as $n^j_i$, by the maximum value in $N$, resulting in a value in the range $[0,1]$ (line 3-5).
Fig.~\ref{heatmap_2} displays the results of Algorithm~\ref{A1} as a heatmap, where inputs $l=60$ and $m=61$ (from March 2016 to March 2021, the takeover month of March 2020 is the 49th month in this case). 
Each cell $(i,j)$ represents the value of $q_i^j$, the per-month normalized block creation rate of the $i^{th}$ BP during the $j^{th}$ month. 
The heatmap provides an intuitive and informative view of the block production activities of the most influential BPs over time.

\noindent \textbf{Observation }
Fig.~\ref{heatmap_2} displays a clear demarcation line in month 0, the takeover month of March 2020, in Steem. 
After the takeover, nearly all the original BPs were removed from the management in Steem, and the positions of the top-20 BPs were firmly controlled by 20 newcomers. 
Most of these newcomers had not created a single block before the takeover.
\textcolor{black}{Specifically, as shown on the left side of Figure 5, before the takeover, the committee was mainly controlled by 27 of the top 30 BPs in Steem, but in the takeover month, the vast majority of these BPs lost their block creation rights, with all of them losing these rights within two months after the takeover. Of the 20 new BPs taking their places, only one had been involved in block creation before the takeover. Ten of these newcomers secured their seats during the takeover month, and the remaining ten solidified their positions within five months following the event. }
In contrast, we did not observe a clear demarcation line in Hive. 
Most original BPs continued to serve Hive after the takeover, and only a few original BPs stopped their services in month 0, the takeover month of March 2020.
\textcolor{black}{Specifically, as depicted on the right side of Figure 5, out of the top 27 BPs in Hive, only two stopped block production after the takeover. Among the remaining 25, 17 maintained the same level of block production, five reduced their activity, and three increased their output. In the 30 to 45 rank range, six previously active BPs increased their block production intensity after the takeover.}
Furthermore, no newcomers were detected after the takeover in Hive.

\noindent \textbf{Insights }
Our results suggest that, after the takeover, the Steem community experienced a clear division. 
Specifically, most of the original management may have migrated from Steem to Hive, while the TRON founder may have gained firm control over the current Steem. 
This finding highlights the potential impact of centralized control on the decentralization of blockchain networks.

\subsection{Voter-layer Decentrlization}
\label{individual}

In this section, we investigate voter-layer decentralization in blockchain networks.
\textcolor{black}{We first propose two metrics, the $k$-entropy coefficient and the $t$-threshold coefficient, to enable measurements across different blockchains from different perspectives.
We then propose a new measurement algorithm named Voter-layer Decentrlization Quantification approach (VLDQ) (Algorithm~\ref{A2}) to quantitatively analyze and compare voter-layer decentralization between different blockchains.
With the proposed metrics and the VLDQ algorithm, we compare voter-layer decentralization between Steem and Hive.}

\noindent \textbf{$k$-entropy coefficient }
In blockchain networks, \textcolor{black}{the term ``whale'' is commonly used to refer to entities with significant resources, capable of influencing the network's dynamics notably.
Studying these whales, especially the top whales of all time, is critical because their long-term influence provides key insights into the structural dynamics of power within the network. By focusing on these dominant players throughout the entire history of the blockchain, we can identify and analyze enduring trends in network decentralization and the concentration of power.
Inspired by previous work that leverages the Shannon entropy to quantify the degree of decentralization in blockchain networks~\cite{li2020comparison,kwon2019impossibility}, we propose the $k$-entropy coefficient to quantify entropy-based decentralization among the top-$k$ whales of all time on a daily basis.
Specifically, let $o_j$ represent the number of active top-$k$ whales on the $j^{th}$ day, where ``active" refers to those whales that have a non-zero number of blocks created on that day, indicating their participation and influence during that specific period. 
Additionally, let $p_{ij}$ denote the percentage of the $i^{th}$ whale's allocated blocks among the active top-$k$ whales' allocated blocks in total on the $j^{th}$ day.}
We compute the normalized entropy for the $j^{th}$ day as follows:
\begin{equation}
e_j = - \sum_{i=1}^{o_j} \frac{p_{ij} \log_2 p_{ij}}{\log_2 o_j} \label{e2}
\end{equation}

\begin{algorithm}[t]
  \footnotesize
  \SetKwInOut{Input}{Input}
  \SetKwInOut{Output}{Output}

  \Input{Blockchain raw data $\mathbf{D}$, daily stake snapshots $\mathbf{S}$.}
  \Output{$k$-entropy coefficient $e$, $t$-threshold coefficient $f$.}

  $\{B,V,\mathbf{C},\mathbf{E}\} \gets parseOps(\mathbf{D})$\;
  Initialize $\mathbf{A}$ for $V$\;
  \For{each day $d$}
  {
    $R \gets rootVoter(V,\mathbf{E})$\;
   \For{each BP $b_i$ in $B$} 
   {
      $\overline{V} \gets supportiveVoter(R,\mathbf{E})$\;
      $\overline{P} \gets stakePercentage(\overline{V},\mathbf{S})$\;
      \For{each supportive voter $\bar{v}_j$ in $\overline{V}$} 
      {
          $k \gets findIndex(\bar{v}_j,V)$\;
          $\mathbf{A}_{k,d}$ += $\bar{p}_j \mathbf{C}_{i,d}$\;
      }
   }
  } 
  
  $e \gets entropy(\mathbf{A},k)$\;
  $f \gets threshold(\mathbf{A},t)$\;
  
  \caption{Voter-layer Decentralization Quantification}
  \label{A2} 
\end{algorithm}

\noindent \textbf{$t$-threshold coefficient }
Thresholds, such as 33\% or 51\%, are frequently used to understand the circumstances of blockchain security~\cite{nakamoto2008bitcoin}.
For instance, the most well-known 51\% attack refers to the situation where a single entity or a group of entities controls more than 51\% of the total computational power in PoW blockchains.
Another example is the PBFT consensus, which requires at least 66\% of the nodes to be honest to ensure the security of the network~\cite{castro1999practical}.
Therefore, another important perspective of decentralization is the estimation of the minimum number of entities whose joint power exceeds a certain threshold.
To quantify threshold-based decentralization on a daily basis, we propose the $t$-threshold coefficient.
\textcolor{black}{
Specifically, let $t$ represent a certain threshold, $N$ denote the set of all entities, $S$ be any subset of $N$, and $u_{ij}$ represent the proportion of blocks allocated to the $i^{th}$ entity on the $j^{th}$ day.
The coefficient for the $j^{th}$ day is computed as follows:}
\begin{equation}
  f_j = \min \left\{ |S| : S \subseteq N, \sum_{i \in S} u_{ij} > t \right\}
\end{equation}

The $t$-threshold coefficient and the $k$-entropy coefficient provide insights into the potential risks of collusion among entities and the balance and equality of whales' power on the network, respectively. 
A higher $t$-threshold coefficient and a higher $k$-entropy coefficient are expected in a more decentralized blockchain network, where the network is less likely to be controlled by a small number of entities (a low $t$-threshold coefficient) or even a single entity (a $t$-threshold coefficient of 1), and the power is more evenly distributed among whales (a high $k$-entropy coefficient).

\begin{figure}
  \centering
  {
      \includegraphics[width=0.60\columnwidth]{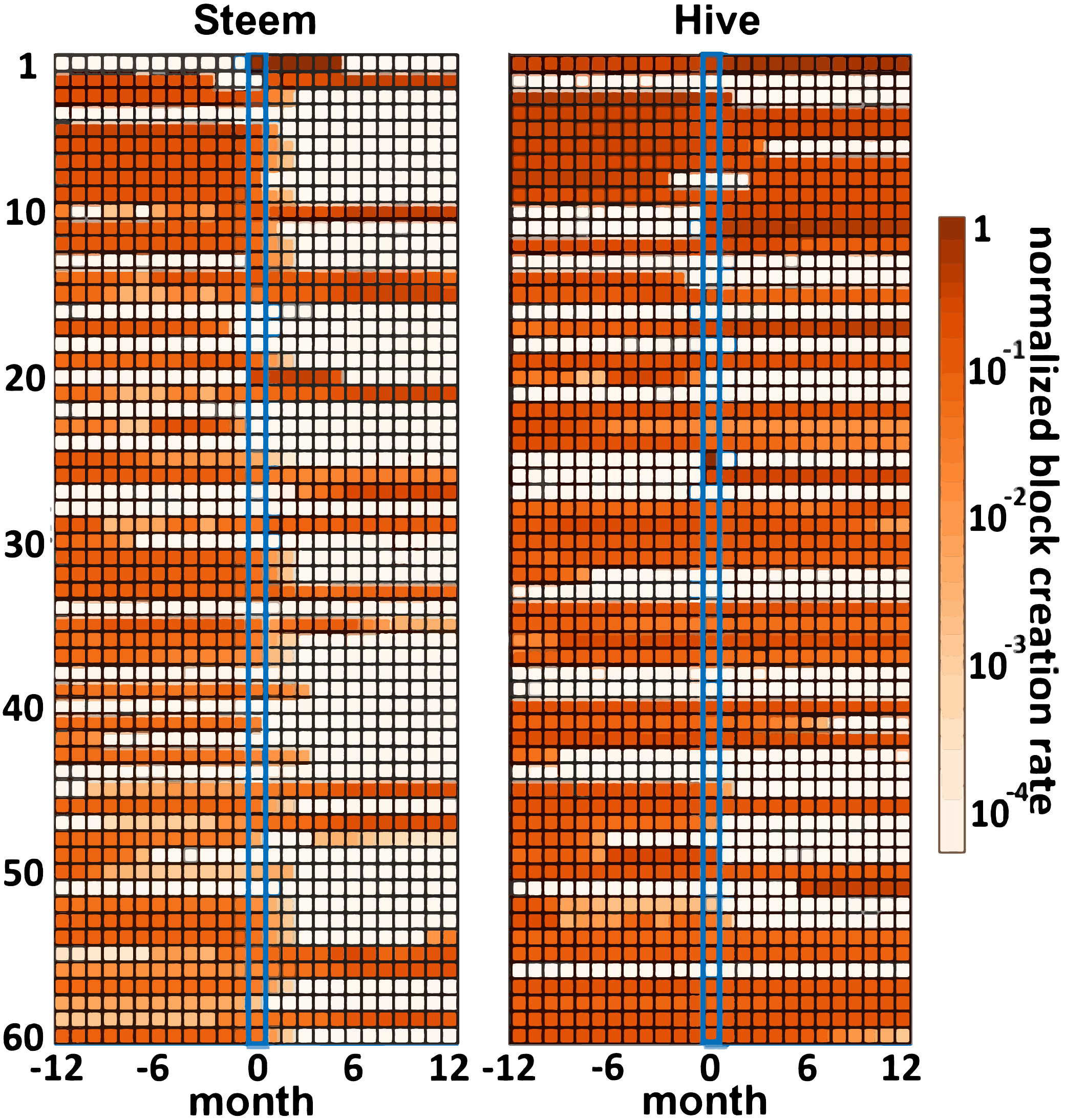}
  }
  \caption {Heatmap of top-60 voters' normalized block creation rates in Steem and Hive during the takeover month (month 0), one year before the takeover (month -12 to -1), one year after the takeover (month 1 to 12)}
  \label{heatmap_4} 
  \end{figure}

\noindent \textbf{Measurement algorithm VLDQ }
There are three main phases in VLDQ:

\begin{itemize}[leftmargin=*]
\item \textbf{Preparation:} 
The algorithm first employs the $parseOps()$ function to extract various types of operations, such as \textit{producer\_reward}, \textit{witness\_vote}, and \textit{witness\_proxy}, from the raw data. 
The resultant output includes a set of block producers ($B$) who have produced at least one block, a set of voters ($V$) who have either voted for BP candidates or have set proxies, the daily BP-block creation snapshots $\mathbf{C}$, and the daily election snapshots $\mathbf{E}$ (line 1). 
Subsequently, it initializes the number of blocks allocated from BPs to the voters in $V$ as $\mathbf{A}$ (line 2).


\item \textbf{Block allocation:} 
For each day $d$, given that voters can set proxies, the algorithm first determines the root voter (the end of a chain of proxies, if any) for each voter in $V$, based on proxy information in $\mathbf{E}$, resulting in $R$ (line 4). Subsequently, for each BP $b_i$ on day $d$, the algorithm performs the following three steps: 
\begin{enumerate}
\item Leveraging the root voters $R$ and election snapshots $\mathbf{E}$, the algorithm identifies all supportive voters $\overline{V}$ who have directly voted for $b_i$ or through their proxy (line 6);
\item Utilizing $\overline{V}$ and $\mathbf{S}$, it calculates the proportion of stake that each supportive voter contributes to the total stake of their group, denoted as $\overline{P}$ (line 7);
\item Depending on $\overline{P}$, $V$, $\overline{V}$, and $\mathbf{B}$, blocks produced by $b_i$ are allocated to supportive voters proportionate to their stake contributions to $b_i$ (line 8-11).
\end{enumerate}


\item \textbf{Decentralization quantification:} 
Finally, based on the metrics proposed in Section~\ref{metrics}, the algorithm computes the $k$-entropy coefficient and the $t$-threshold coefficient, respectively (line 14-15).
\end{itemize}

\begin{figure}
  \centering
  {
      \includegraphics[width=0.70\columnwidth]{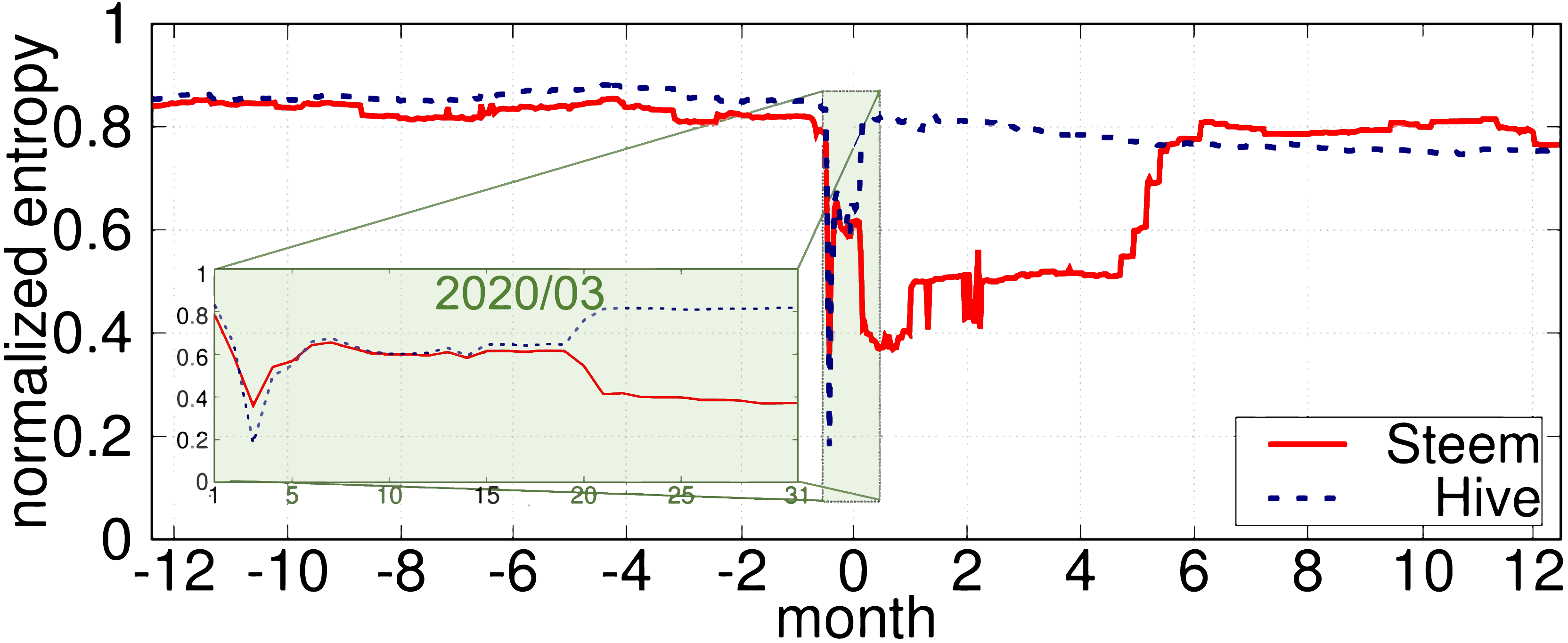}
  }
  \caption {\small The $100$-entropy coefficient computed on a daily basis in Steem and Hive before and after the takeover month (month 0), \textcolor{black}{where the discrepancies before the fork are attributed to the differences in the composition of the top-100 whales of all time between Steem and Hive. An inserted plot details the coefficient changes during the takeover month, showing a sharp drop and subsequent rise in the coefficients around the takeover day, culminating in the divergence of Steem and Hive on the day of Marth 18}}
  \label{NE_2} 
  \end{figure}

\begin{figure}
    \centering
    {
        \includegraphics[width=0.70\columnwidth]{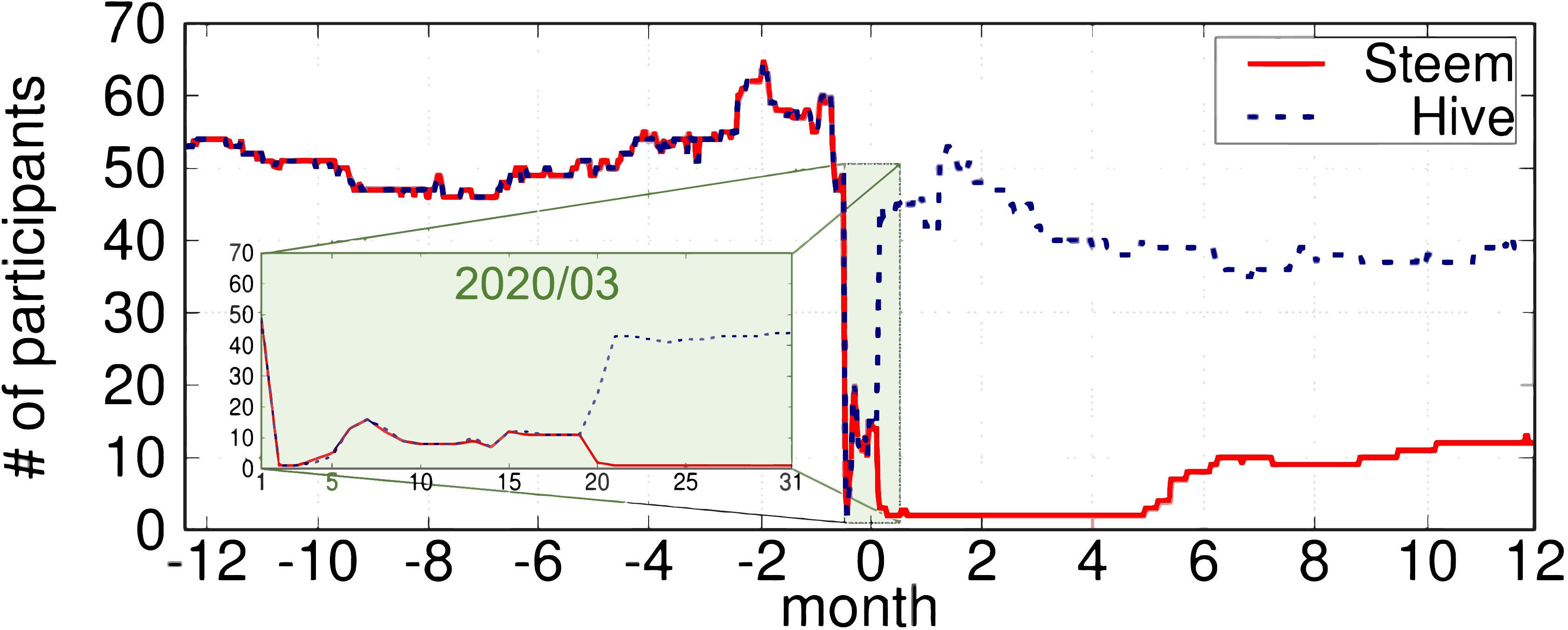}
    }
    \caption {\small The 50\%-threshold coefficient computed on a daily basis in Steem and Hive before and after the takeover month (month 0). \textcolor{black}{An inset plot illustrates the daily fluctuations in the coefficients throughout the takeover month, documenting a pronounced decline followed by a recovery leading up to the day when Steem and Hive split on the day of Marth 18.}}
    \label{MF_2} 
    \end{figure}

We utilize the VLDQ algorithm to allocate blocks from BPs to voters in both Steem and Hive. 
The results are presented in Fig.~\ref{heatmap_4}, Fig.~\ref{NE_2}, and Fig.~\ref{MF_2}, respectively.

\noindent \textbf{Observation }
Firstly, by slightly modifying the visualization methodology presented in Section~\ref{pool}, Fig.~\ref{heatmap_4} displays a heatmap measuring the normalized block creation rates of the top-60 voters in Steem and Hive, instead of the top-60 BPs, after allocating blocks from BPs to voters. 
Similar to the results displayed in Fig.~\ref{heatmap_2}, we observe a clearer demarcation line in Steem. 
Specifically, we note that most whales (i.e., top voters) left Steem within one or two months after the takeover month (month 0). 
In contrast, only a small percentage of whales stopped electing BPs in Hive. 
It is worth noting that both the $1^{st}$ (\textit{@steemit}) and $15^{th}$ voters in Steem are accounts restricted by Fork 0.22.2. 
These two accounts started to receive allocated blocks in the takeover month 0 and stopped receiving allocated blocks in month 5. 
It is also surprising to note that \textit{@steemit} won first place by only taking part in the BP election for around six months.

Fig.~\ref{NE_2} displays the $100$-entropy coefficient (i.e., $k$-entropy coefficient with $k=100$) for Steem and Hive during the period from March 2019 to March 2021. 
We observe a sharp drop in the $100$-entropy coefficient from 0.8+ to 0.4- around the takeover day of March 2, 2020. Subsequently, the $100$-entropy coefficient quickly recovered to reach 0.6. 
On the Hive fork day of March 20, 2020, we observe an upsurge in Hive's $100$-entropy coefficient, while Steem's $100$-entropy coefficient experienced an abrupt reduction. 
However, after approximately five months, the $100$-entropy coefficients of both Steem and Hive returned to comparable levels.

Fig.~\ref{MF_2} presents the $50\%$-threshold coefficient (i.e., $t$-threshold coefficient with $t=50\%$) for Steem and Hive during the period from March 2019 to March 2021. 
Similar to the trends shown in Fig.~\ref{NE_2}, we observe a sharp drop in the $50\%$-threshold coefficient around the takeover day of March 2, 2020, from 49 to 2. 
Subsequently, the coefficient recovered to around 13. 
However, on the Hive fork day of March 20, 2020, Steem's $50\%$-threshold coefficient dropped back to 5 from 14, while Hive's $50\%$-threshold coefficient increased to 29 from 14. 
Compared with the changes in $100$-entropy coefficients shown in Fig.~\ref{NE_2}, we observe that the changes in $50\%$-threshold coefficients are quite different. 
Specifically, we note that the gap between Hive's and Steem's $50\%$-threshold coefficients remains significant, even one year after the takeover.


\noindent \textbf{Insights }
\textcolor{black}{Our results reveal that Hive tends to be more decentralized than Steem was after the takeover, yet it does not reach the level of decentralization that Steem maintained prior to the takeover.}
Specifically, we observe that the voter-layer decentralization of Hive is higher than that of Steem. 
However, both Steem and Hive experienced a drop in decentralization after the takeover, as evidenced by the decrease in the coefficients in Fig.~\ref{NE_2} and Fig.~\ref{MF_2}. 
Moreover, our findings demonstrate the long-term damage to the decentralization of both Steem and Hive after the takeover, which may suggest that no one won the battle from this perspective.

It is worth noting that the decentralization of a blockchain network is a complex and multifaceted concept that cannot be fully captured by a single metric. 
Our study focuses on the BP-layer and voter-layer decentralization of Steem and Hive, which is only one aspect of the overall decentralization of these two blockchains. 
Future research could explore other dimensions of decentralization regarding Web 3.0, such as the degree of censorship resistance, and the level of community participation.

\section{Unveiling the Takeover Strategies}
\label{sec5}

In this section, after empirically illuminating the impact of blockchain network takeovers on decentralization, 
we move one step forward to develop a fine-grained understanding of the underlying mechanics of takeover strategies.
Specifically, we first extract distinct voting strategies performed by voters and statistically analyze the voting strategies.
We then propose heuristic methods to identify anomalous voters and conduct clustering analyses on voter behaviors.
Finally, we suggest potential mitigation strategies that can be applied to fortify future Web 3.0 network against takeovers.

\begin{figure}
  \centering
  {
      \includegraphics[width=0.70\columnwidth]{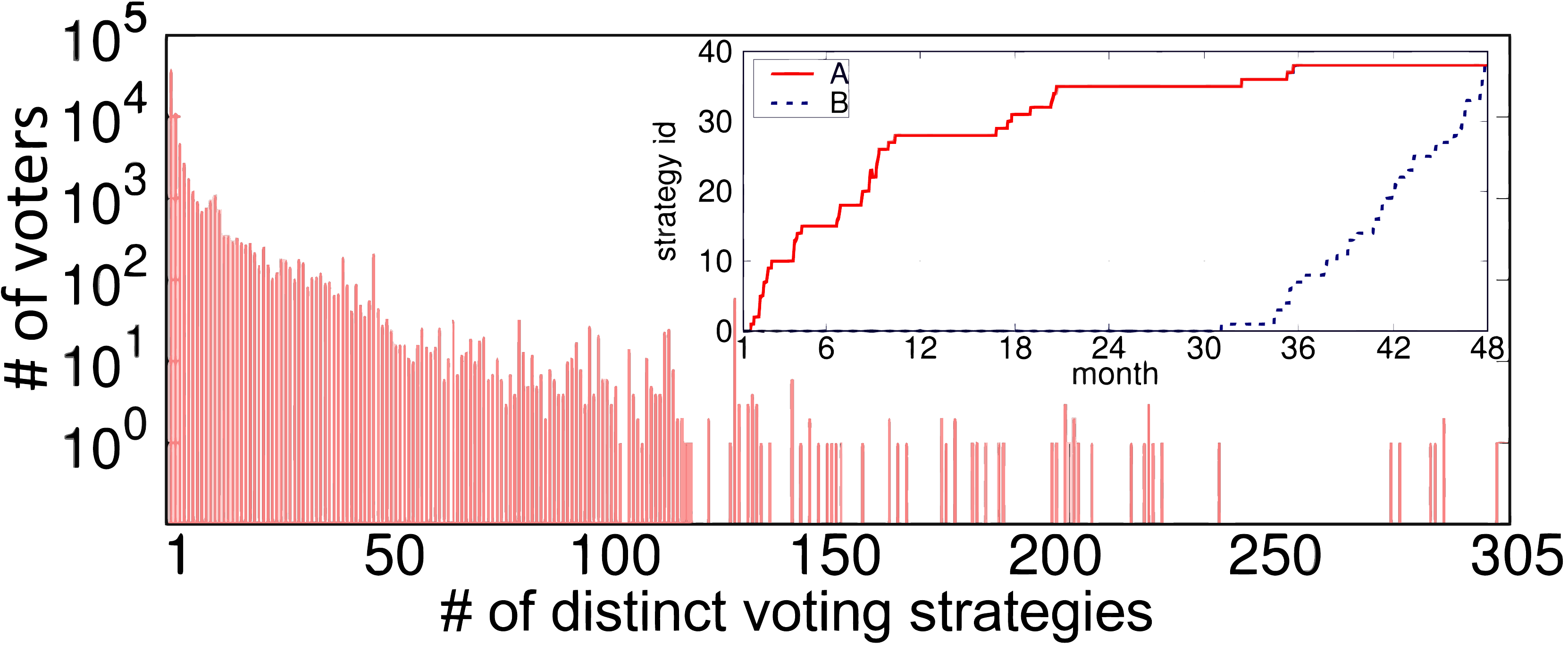}
  }
  \caption {\small Statistics for the number of distinct voting strategies employed by voters}
  \label{cluster_1} 
\end{figure}


\subsection{Extracting Distinct Voting Strategies}
\label{voting_strategy}

\noindent \textbf{Methodology }
In Steem and Hive, each voter may cast votes for up to 30 distinct BPs. 
In this paper, we define a voting strategy of a voter as the set of BPs they have voted for. 
Our definition of distinct voting strategies is strict, such that two voting strategies that differ in a single BP are considered distinct. 
Therefore, we extract a set of all the distinct voting strategies performed by voters, which includes the various combinations of BPs that any voter has chosen in the history of BP election.

\noindent \textbf{Observation }
In Fig.~\ref{cluster_1}, we investigate the number of distinct voting strategies employed by each of the 73,430 voters that have participated in BP election. 
The results show that 38,455 (51\%) voters have employed only a single voting strategy, while 11,320 (15\%) voters have employed only two distinct voting strategies. 
However, we also observe that some voters frequently change their voting strategies, with the highest number of distinct voting strategies employed by a single voter being 302.
Fig.~\ref{cluster_1} also illustrates the existence of different patterns of strategy changes. 
The figure presents an example involving two voters, A and B, who both employed 38 distinct voting strategies in total. 
Voter A participated in the election very early and did not frequently change the voting strategy. 
In contrast, voter B participated in the election quite late and frequently changed the voting strategy since month 34.

\begin{figure}
  \centering
  {
      \includegraphics[width=0.70\columnwidth]{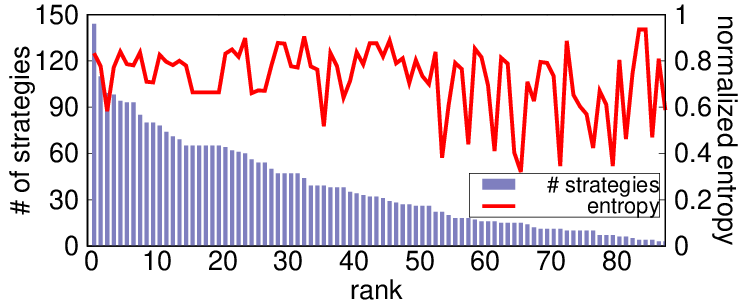}
  }
  \caption {\small Statistics for the number of distinct voting strategies employed by top-100 voters}
  \label{sec6_1_03} 
\end{figure}

In Figure~\ref{sec6_1_03}, we analyze the number of distinct voting strategies employed by the top-100 voters. 
These voters are selected in the same way as in Figure~\ref{heatmap_4}, but are reranked based on the number of distinct voting strategies they have employed. 
We find that only 5 of them have never changed their voting strategies. 
In contrast, 77 of them have employed more than 10 distinct voting strategies, and 28 of them have employed more than 50 distinct voting strategies.
To investigate the duration of each voting strategy, we compute the normalized entropy across the number of days spent on different voting strategies for each of the selected voters. 
The results show that most of the selected voters who have explored more voting strategies tend to have higher normalized entropy, which indicates that they have spent a similar amount of time on each voting strategy.

\noindent \textbf{Insights }
Our findings suggest that the majority of the top voters are not loyal enough to any voting strategy. 
We can recognize a trend that a voter who has employed more voting strategies may distribute the days more evenly to different voting strategies. 
This finding is particularly relevant in the context of the takeover, as it suggests that the top voters may not have been committed to supporting the same group of BPs.
For decentralization, being less loyal may be a better choice. 
Instead of sticking to a fixed set of BPs, a more dynamic change of voting strategies could potentially promote the increment of the level of decentralization by excluding outdated BPs. 

\subsection{Identify Anomalous Voters}
\label{anomalous_voter}
\noindent \textbf{Methodology }
In the next step, we propose heuristic methods to leverage the set of distinct voting strategies to identify anomalous voters that may have played decisive roles in the takeover. 
We apply three detection criteria to identify anomalous voters:
\begin{enumerate}
\item Inactivity Period: 
The voter has refrained from participating in BP elections, neither casting votes nor setting proxies, for a duration of at least one year.
\item Sudden Activity Resumption: 
The voter resumes participation in BP elections abruptly after a dormancy period of at least one year.
\item Unique Voting Strategy: 
The voting strategy used by the voter upon reappearing on day $d$ should be unique and not a member of the set of distinct strategies that have been used from day 1 through $d-1$.
\end{enumerate}



\begin{figure}
  \centering
  {
    \includegraphics[width=0.70\columnwidth]{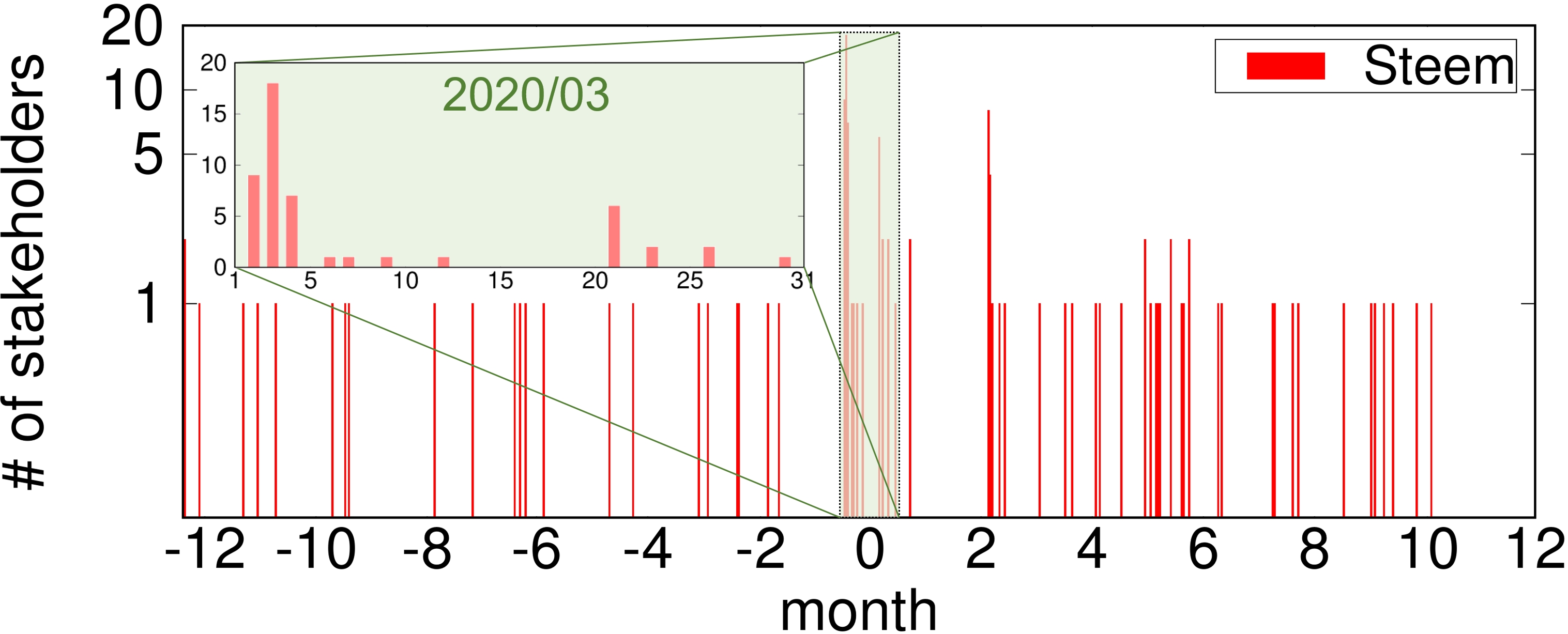}
  }
  \caption {\small The per day number of anomalous voters in Steem during two years before and after the takeover month (month 0)}
  \label{anomalous_s} 
\end{figure}

\begin{figure}
  \centering
  {
      \includegraphics[width=0.70\columnwidth]{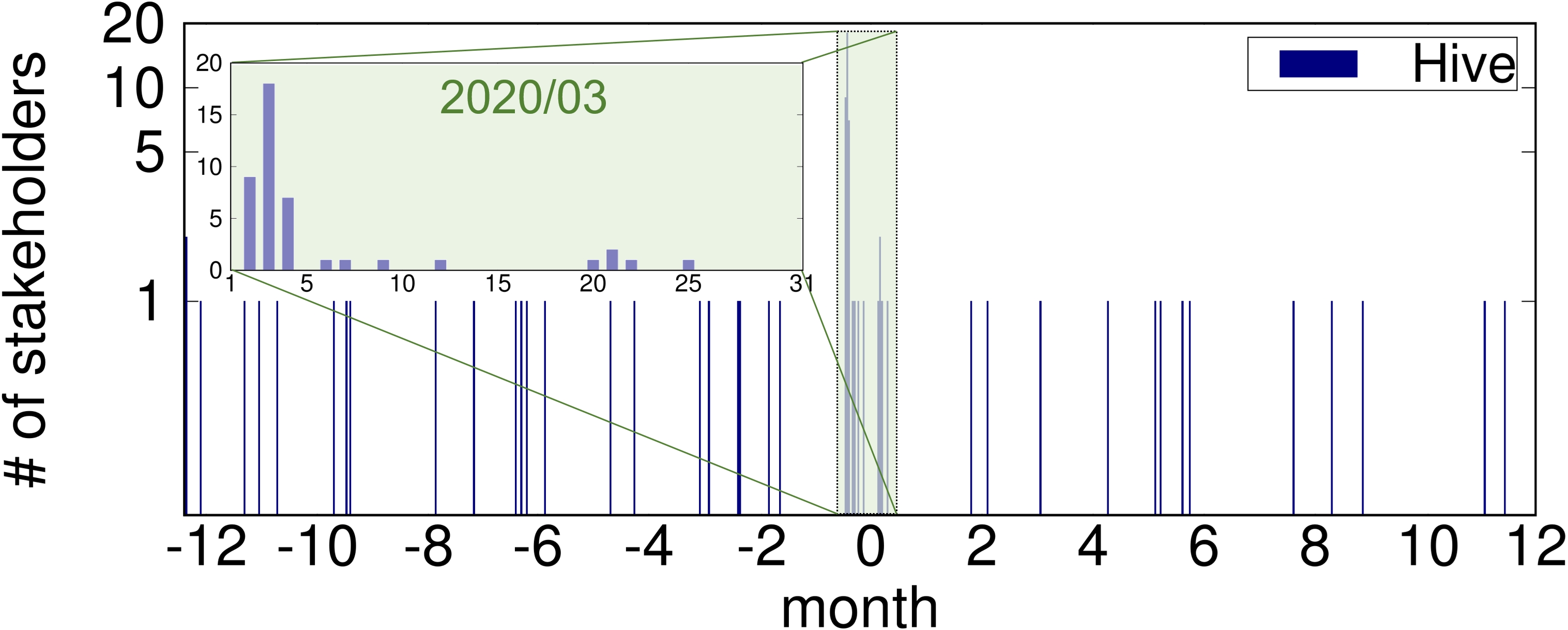}
  }
  \caption {\small The per day number of anomalous voters in Hive during two years before and after the takeover month (month 0)}
  \label{anomalous_h} 
\end{figure}


\noindent \textbf{Observation }
In Fig.~\ref{anomalous_s}, we detect the anomalous voters in Steem who started to participate in the BP election with a distinct voting strategy around the takeover day. 
The results indicate that the anomalous voters satisfying the prescribed criteria could not be commonly detected before the takeover month. 
However, on the takeover day of March 2, 2020, we could detect nine anomalous voters that match the prescribed criteria. 
After the takeover day and even the Hive fork day, we could detect two surges of such anomalous voters.

In Fig.~\ref{anomalous_h}, we detect the anomalous voters in Hive that satisfy the aforementioned conditions. 
Compared with the results in Steem shown in Fig.\ref{anomalous_s}, we observe that after the Hive fork day on March 20, 2020, there is no surge of such anomalous voters in Hive.

\noindent \textbf{Insights }
Our results suggest that the nine anomalous voters detected on March 2, 2020, may have played a decisive role in the takeover process.
We will further investigate these voters in the next section to gain a better understanding of their behavior.
Furthermore, our findings suggest that the Hive fork may have been successful in preventing the emergence of such anomalous voters, which may demonstrate that a hard network fork can be an effective mitigation strategy to prevent similar takeovers in the future. 

\subsection{Clustering Analyses on Voter Behaviors}

\noindent \textbf{Methodology. }
We propose a heuristic approach for identifying clusters of voters, defined as groups of voters who have exhibited identical changes in their voting strategies over a specific period of time. 
Our heuristic method, named Cluster Identification Method (CIM), involves the following steps:
\begin{enumerate}
  \item Labeling Voting Strategies: In the first step, CIM assigns unique identifiers (serial numbers) to each distinct voting strategy employed by the selected set of voters.

  \item Tracing Voting Behavior: CIM then systematically tracks the daily voting strategies of each selected voter, creating a chronological record of voting behavior alterations.

  \item Grouping by Similar Changes: On a day-to-day basis, CIM isolates clusters of voters who have made identical modifications to their voting strategies.

  \item Identifying Significant Clusters: In the final step, CIM validates these groups. If a group of voters has exhibited synchronized voting strategy changes at least $l$ times, the group is recognized as a significant cluster.
\end{enumerate}
By following these heuristic steps, CIM allows for a more efficient and streamlined process to detect clusters of voters based on their shared voting strategy alterations. Note that the choice of $l$, which defines the minimum occurrences of synchronized strategy changes required for a group to be considered a cluster, can be adjusted based on specific requirements or thresholds.
In our analysis, we assign the strategy with $sn=0$ to indicate the case where a voter casts no direct or indirect votes. For example, the notation $A=[130,(49,353)]$ indicates that voter A changed their voting strategy from $sn=49$ to $sn=353$ on day 130.



\noindent \textbf{Observation }
We apply CIM to the top-100 voters selected in the same way as in Fig.~\ref{heatmap_4} and Fig.~\ref{sec6_1_03}. 
Surprisingly, the results reveal no group of voters that have simultaneously changed their voting strategies only once, which may suggest that such a coincidence is rare.
Instead, we find six non-overlapping clusters of voters among these top-100 voters. 
Voters in each cluster have simultaneously changed their voting strategies at least 10 times.

\begin{figure*}
  \centering
  \subfigure[{\small The first cluster}]
  {
     \label{sec6_2_01}
      \includegraphics[width=0.48\columnwidth]{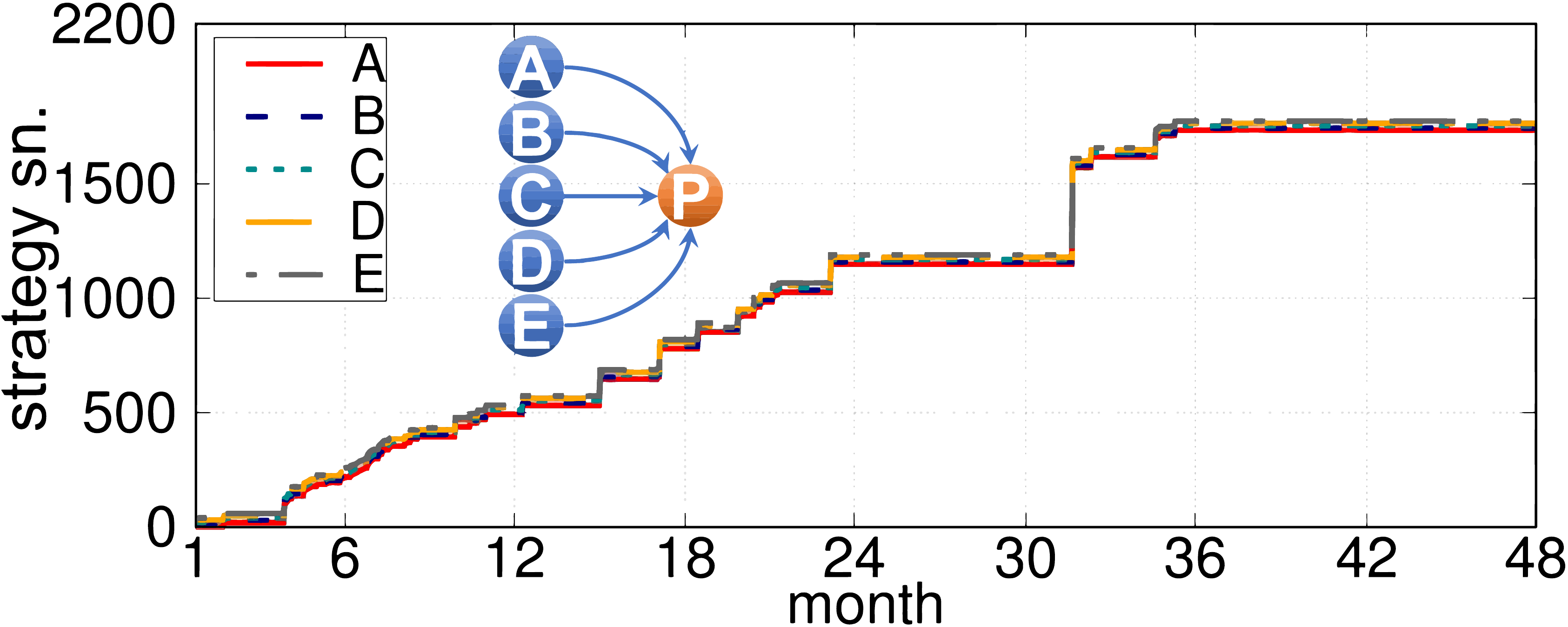}
  }
  \subfigure[{\small The second cluster}]
  {
    \label{sec6_2_02}
      \includegraphics[width=0.48\columnwidth]{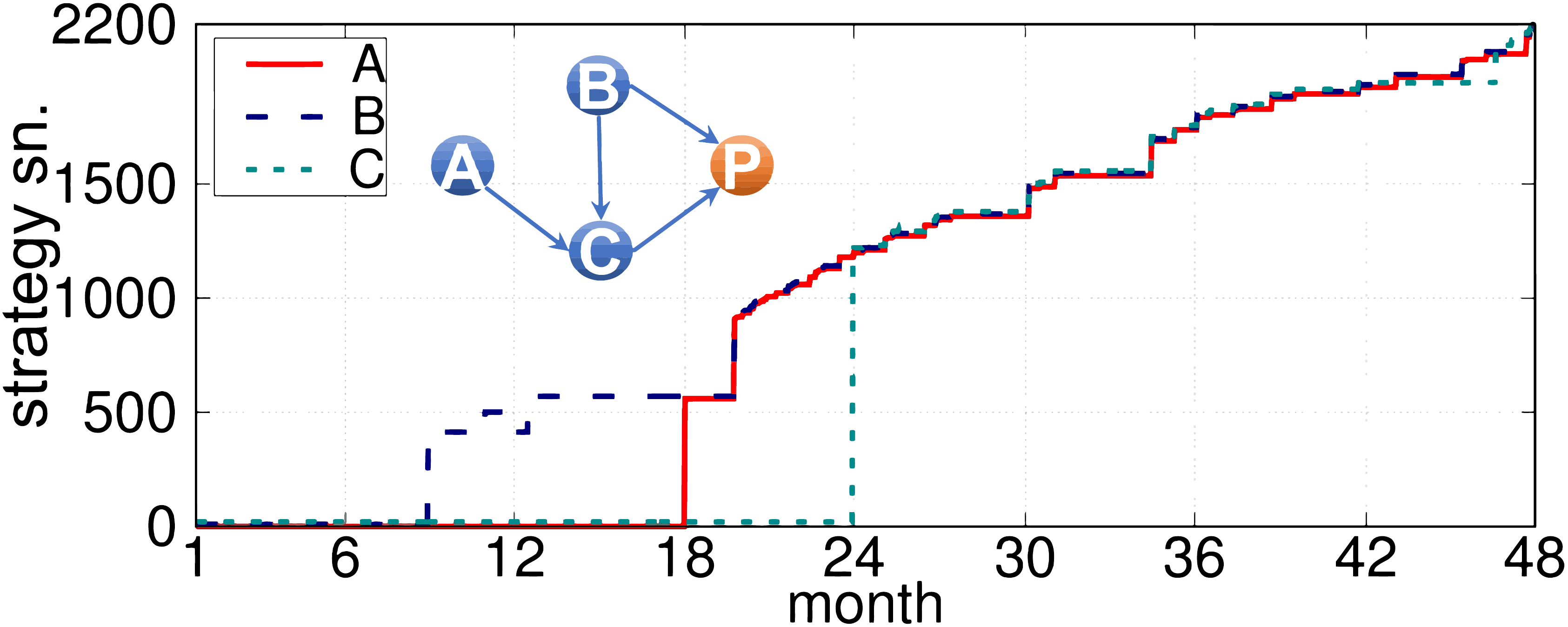}
  }
  \subfigure[{\small The third cluster}]
  {
     \label{sec6_2_03}
    \includegraphics[width=0.48\columnwidth]{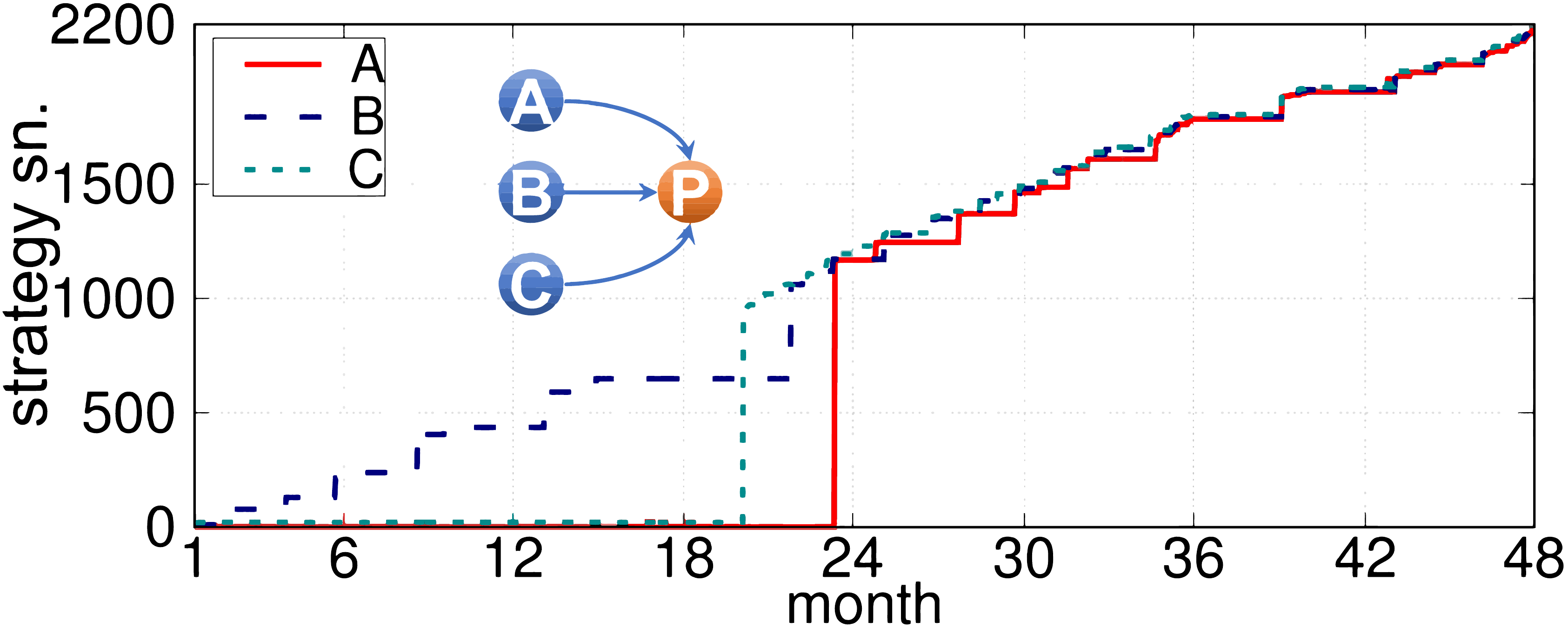}
  }
  \subfigure[{\small The fourth cluster}]
  {
     \label{sec6_2_04}
    \includegraphics[width=0.48\columnwidth]{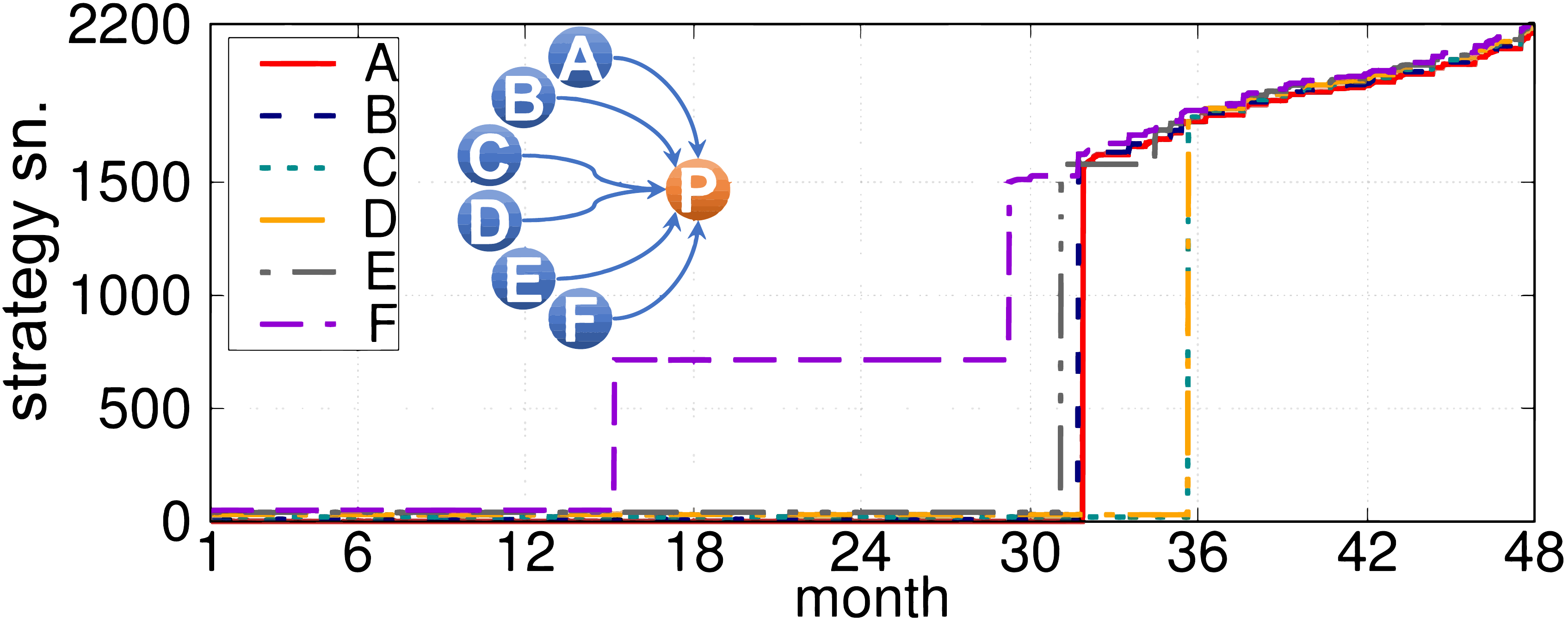}
  }
  \subfigure[{\small The fifth cluster}]
  {
    \label{sec6_2_05}
      \includegraphics[width=0.48\columnwidth]{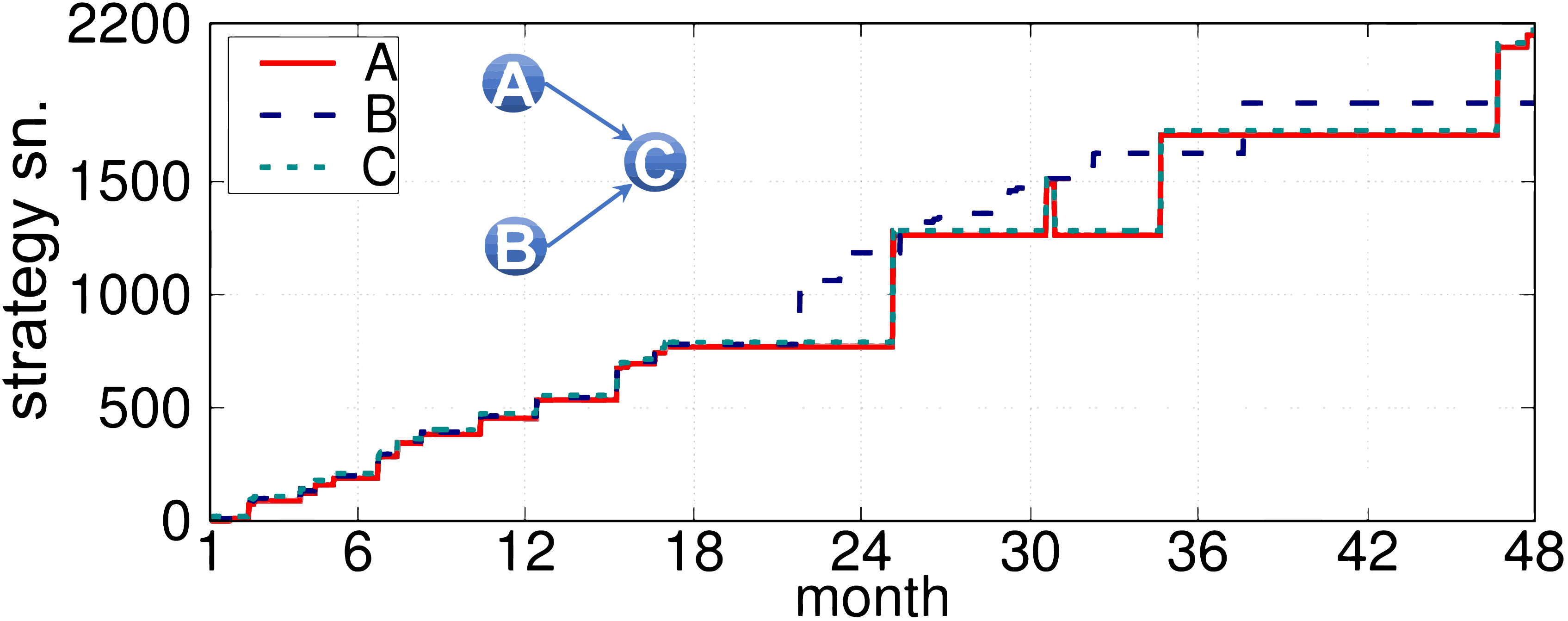}
  }
  \subfigure[{\small The sixth cluster}]
  {
     \label{sec6_2_06}
    \includegraphics[width=0.48\columnwidth]{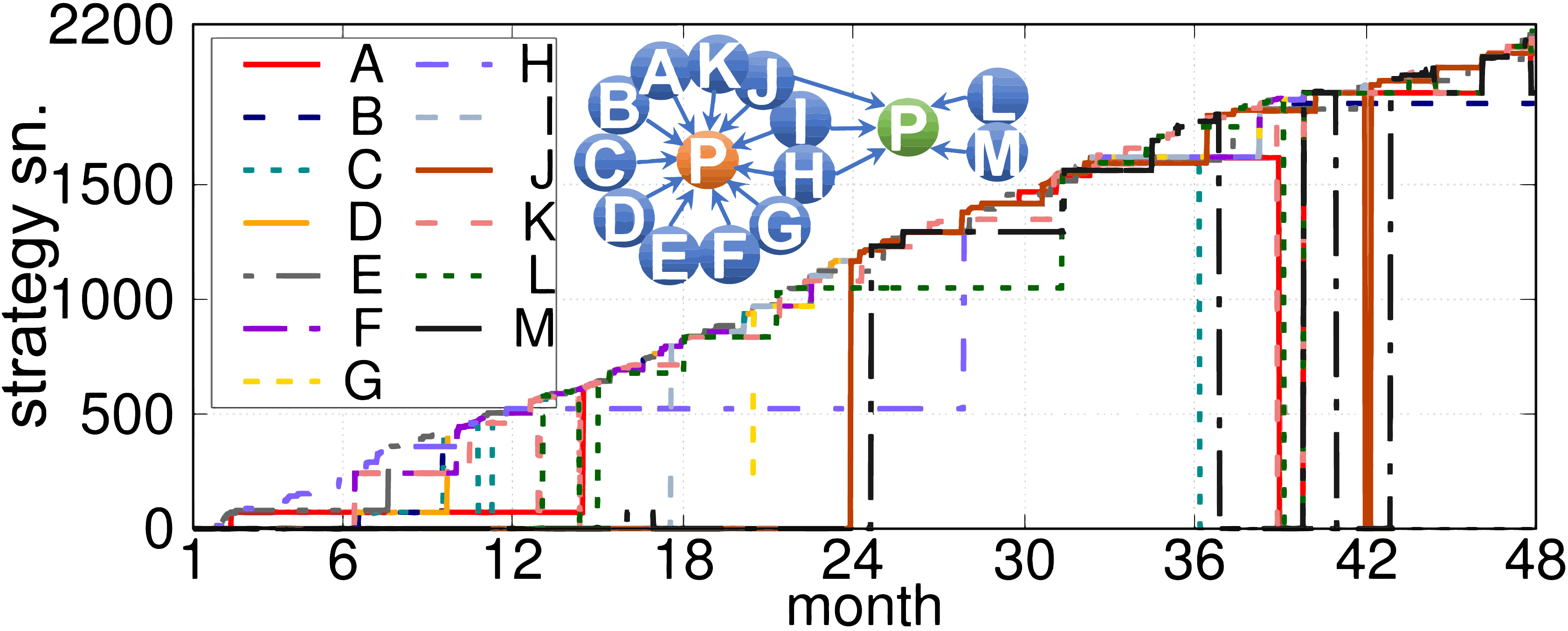}
  }
  \caption{Strategy change patterns and proxying relationships in six clusters of voters that have simultaneously changed their voting strategies at least 10 times}
  \label{cluster_4}
\end{figure*}

Fig.~\ref{cluster_4} illustrates the patterns of strategy changes for the six clusters, as well as the proxying relationships among members of each cluster.
The first cluster consists of five members who changed their voting strategies 64 times at exactly the same pace. 
All five members have set the same voter outside the cluster as their proxy during their lifetimes, which we refer to as the outward proxying relationship.
In the second cluster, three members changed their voting strategies 25 times at exactly the same pace. 
We observe a combination of inward and outward proxying relationships. 
Specifically, two members, A and B, have set member C as their proxy, while B and C have also set another voter outside the cluster as their proxy.
The third and fourth clusters reveal outward proxying relationships among three and six members, respectively. 
In the fifth cluster, we observe a pure inward proxying relationship. 
Two members of the fifth cluster, A and B, have set the rest of the members as their proxy. 
More concretely, A has stuck to the proxy during its lifetime, while B canceled the proxy in month 22 and directly cast votes to BPs from then on.
In the sixth cluster, we observe more complicated outward proxying relationships. 
This large cluster consists of thirteen members who joined the cluster at different time points, and some have left or even stopped voting (i.e., changed strategy to $sn=0$). 
The thirteen members form two overlapped outward proxying relationships. 
Specifically, eleven members, A to K, have set one voter outside the cluster as their proxy, while three of these eleven members (J, I, and H) and the rest two members (L and M) have ever set another voter outside the cluster as their proxy.

\begin{figure}
  \centering
  {
      \includegraphics[width=0.70\columnwidth]{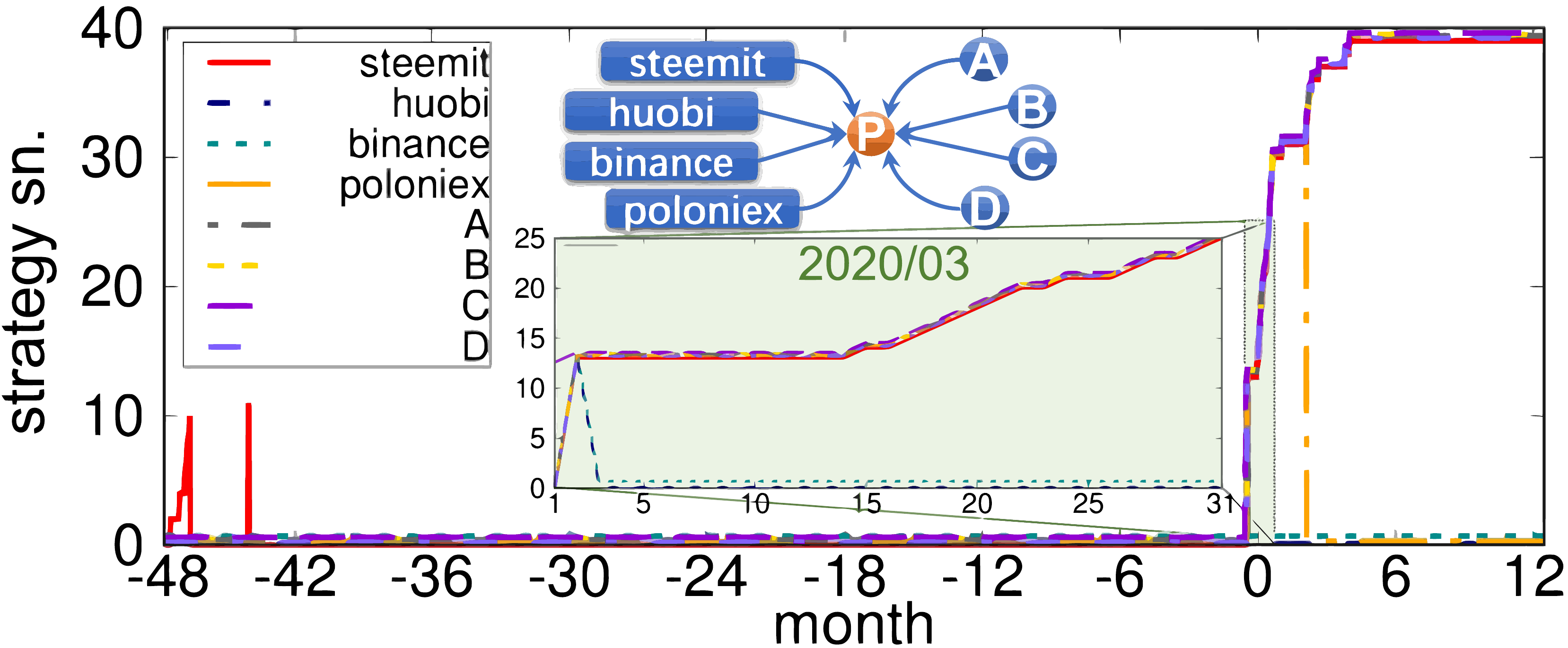}
  }
  \caption {\small Strategy change patterns and proxying relationships in the cluster of eight voters detected on the takeover day}
  \label{takeover_cluster} 
\end{figure}

Next, we apply the voter clustering heuristic CIM to the nine anomalous voters detected on the takeover day, as discussed in Section~\ref{anomalous_voter}. Fig.~\ref{takeover_cluster} shows the results of this analysis.
Among the nine anomalous voters, we detect a cluster of eight voters who have simultaneously changed their voting strategies at least once. 
This cluster includes three exchanges and three accounts restricted by the Fork 0.22.2 (e.g., \textit{@steemit}). 
Among them, \textit{@steemit} employed 11 distinct voting strategies before July 2016 but has not been involved in the election since August 2016. 
The rest of the accounts have never participated in the election until the takeover day.
On the takeover day, all eight voters suddenly appeared in the election and employed the same voting strategy $sn=13$, which had never been seen before. 
Immediately after the takeover, Binance and Huobi left the election by setting their voting strategy back to $sn=0$. 
Poloniex left the election after about two and a half months. 
During the entire process, all these voters set the same voter outside the cluster as their proxy.

\noindent \textbf{Insights }
Our main insight is that the takeover was implemented by well-known voters using well-known stakes, rather than by unknown voters using unknown stakes. 
Most users in Steem were aware that the accounts belonged to exchanges or Steemit Inc. and held large amounts of stake, either from users of exchanges or from pre-mined tokens. 
However, since both the exchanges and Steemit Inc. were widely considered to be `neutral', the takeover may have been considered to be only theoretically possible.
This phenomenon is not unique to Steem. 
In fact, many public blockchains have pre-mined tokens or founder's rewards. 
For example, in EOSIO, another prominent DPoS blockchain, many tokens are currently controlled by exchanges, including those involved in the Steem takeover. 
Additionally, as is well-known, many computational resources in public blockchains such as Bitcoin~\cite{zheng2023bshunter} and Ethereum~\cite{zheng2020xblock} are currently controlled by pools.
There have been opinions that these `neutral' actors may never choose to misuse their power in practice, from the perspectives of incentive compatibility or the value of reputation~\cite{zeng2021characterizing}. 
However, the TRON-Steem takeover incident demonstrates that `neutral' actors may misuse their power in practice, rising as a threat to the security of Web 3.0.


\subsection{Potential Mitigation Strategies}
\label{sec:mitigation}
As illuminated by the analyses in Section~\ref{anomalous_voter}, a hard fork is a practical strategy to resist a hostile takeover. 
However, as discussed in Section~\ref{sec4}, a hard fork may not be the first choice due to potential drawbacks, such as a loss of decentralization in blockchain networks and a loss of trust in the community, which may lead to developers and investors losing confidence in the blockchain and leaving the community.

One potential strategy to suppress the power of hostile voters is to reduce the weight of their votes. 
\textcolor{black}{This can be achieved by limiting the number of votes that a voter can cast. }
Currently, EOSIO, Steem, and TRON allow each voter to cast 30 stake-weighted votes, enabling a single hostile voter to support up to 30 BPs. 
Intuitively, if the number of votes that a voter can cast is limited to 1, the power of hostile voters will be significantly reduced. 
\textcolor{black}{This reduction in voting power would make it more challenging for a small group of voters to exert disproportionate control over the network.}
Another potential strategy is to implement a delay period for votes to become effective. 
This approach ensures that malicious votes do not immediately take effect, affording the community time to respond. 
By implementing such a delay period, the community can detect and respond to hostile takeover attempts before they can cause significant damage.
\textcolor{black}{This delay period would act as a buffer, giving stakeholders the opportunity to mobilize and counteract any malicious actions.}

\textcolor{black}{Additionally, introducing weighted voting based on voter reputation or historical behavior could further enhance the network's resilience against hostile takeovers. By assigning more weight to votes from reputable and long-standing community members, the influence of new or potentially malicious voters could be mitigated. This approach encourages positive participation and long-term commitment to the network.}
\textcolor{black}{Furthermore, transparent governance practices and enhanced community engagement are crucial in preventing hostile takeovers. Regular communication and updates from network developers and administrators can build trust and ensure that the community remains vigilant against potential threats. Encouraging active participation in governance and decision-making processes helps to create a more decentralized and resilient network.}

\textcolor{black}{In conclusion, while a hard fork is a viable option, it is not without its drawbacks. The strategies discussed above offer alternative and complementary approaches to mitigate the risk of hostile takeovers. These measures can help maintain the integrity and decentralization of blockchain networks, fostering a more secure and trustworthy environment for all participants.}



\section{Related work}
\label{sec6}
The majority of recent studies analyzing decentralization of blockchain networks have centered around Bitcoin~\cite{gervais2014bitcoin,wang2020measurement,miller2015discovering,zheng2023blockchain}.
In~\cite{gervais2014bitcoin}, the centralization of Bitcoin in Bitcoin Web Wallets, Protocol Maintenance, and BlockChain Forks was analyzed, and possible ways to enhance decentralization were proposed. 
Through the research of a small fraction of the network, authors in~\cite{miller2015discovering} found that nodes connected with major mining pools occupy higher mining capacity. 
From the perspective of mining pools, authors in~\cite{wang2020measurement} tracked more than 1.56 million blocks (including about 257 million historical transactions) and found that a few mining pools were controlling and will keep controlling most of the computing resources of the Bitcoin network.
In addition to Bitcoin, the level of decentralization in Steem has been examined in recent studies~\cite{li2019incentivized}. The results indicated a comparatively low degree of decentralization within the network.

Recent works have compared decentralization across multiple blockchain networks~\cite{gencer2018decentralization,kwon2019impossibility,wu2019information,zeng2021characterizing,li2020comparison,li2023cross,liu2023demonitor}.
The work in~\cite{gencer2018decentralization} relied on a new measurement technology to obtain application layer information and focused on the Falcon network. 
The work in~\cite{wu2019information} investigated Bitcoin and Ethereum and suggested that the degree of decentralization in Bitcoin was higher than that in Ethereum.
Authors in~\cite{kwon2019impossibility,li2020comparison} analyzed the degree of decentralization in both Bitcoin and Steem. However, without fixing the missing system parameter $\lambda$ and reconstructing both historical stake snapshots and election snapshots for Steem, they only measured the temporal level of decentralization in Steem.
The work in~\cite{zeng2021characterizing} measured the degree of decentralization in Ethereum at the level of mining pool participants. The results indicated that decentralization measured at a deeper level (i.e., across participants of mining pools in Ethereum) could be quite different from that measured across mining pools.
The work in~\cite{li2023cross} investigated Ethereum and Steem using a novel two-level comparison framework, suggesting that the individual-level decentralization in Steem may present centralization risks.


Among the recent studies on the Steem blockchain~\cite{ba2022fork,guidi2022assessment,guidi2021analysis,li2023liquid,li2023hard,li2023characterizing,jeong2020centralized}, the most relevant works to ours are~\cite{jeong2020centralized}, ~\cite{ba2022fork} and~\cite{li2023hard}.
\textcolor{black}{The work in~\cite{jeong2020centralized} was the first to analyze the TRON-Steem takeover incident and found the optimal number of votes per account that minimizes takeover risks.}
Authors in~\cite{ba2022fork} analyzed user migration from Steem to Hive following the TRON-Steem takeover incident. 
The study presented in~\cite{li2023hard} examines the security aspects of the coin-based voting governance system inherent in DPoS blockchains and proposes potential enhancements to bolster the resilience of any blockchain utilizing this governance model against takeovers.
To the best of our knowledge, our work in this paper presents the first large-scale longitudinal study that empirically highlights the severe threat that blockchain network takeovers pose to the decentralization principle of Web 3.0. 
We unveil the underlying mechanics of takeover strategies employed in the Tron-Steem incident and suggest potential mitigation strategies, which contribute to the enhanced resistance of Web 3.0 networks against similar threats in the future.

\section{Conclusion}
\label{sec7}

In this paper, we present a large-scale data-driven analysis of the TRON-Steem takeover incident, the first de facto blockchain network takeover in Web 3.0. 
To empirically demonstrate the impact of blockchain network takeovers on decentralization, we propose a layered measurement model, two metrics, and a measurement algorithm VLDQ. 
We also fix the missing system parameter $\lambda$ and reconstruct the daily stake and election snapshots.
Our results reveal the long-term damage to the decentralization of both Steem and Hive after the takeover, which may demonstrate the severe threat that blockchain network takeovers pose to the decentralization principle of Web 3.0. 
To develop a deeper understanding of the underlying mechanics of takeover strategies, we propose heuristic methods to identify anomalous voters and conduct clustering analyses on voter behaviors. 
Specifically, our findings warn the community about the potential collusion of neutral actors, such as accounts holding pre-mined tokens and exchanges holding users' tokens, to takeover blockchain networks.
Finally, we suggest potential mitigation strategies that can be applied to fortify future Web 3.0 network against takeovers. 
Our findings may be of interest to a broader audience of blockchain and Web 3.0 practitioners and researchers, as they contribute to the enhanced resistance of Web 3.0 networks against similar threats in the future.

\bibliographystyle{ACM-Reference-Format}
\bibliography{main}


\begin{thebibliography}{52}


\ifx \showCODEN    \undefined \def \showCODEN     #1{\unskip}     \fi
\ifx \showDOI      \undefined \def \showDOI       #1{#1}\fi
\ifx \showISBNx    \undefined \def \showISBNx     #1{\unskip}     \fi
\ifx \showISBNxiii \undefined \def \showISBNxiii  #1{\unskip}     \fi
\ifx \showISSN     \undefined \def \showISSN      #1{\unskip}     \fi
\ifx \showLCCN     \undefined \def \showLCCN      #1{\unskip}     \fi
\ifx \shownote     \undefined \def \shownote      #1{#1}          \fi
\ifx \showarticletitle \undefined \def \showarticletitle #1{#1}   \fi
\ifx \showURL      \undefined \def \showURL       {\relax}        \fi
\providecommand\bibfield[2]{#2}
\providecommand\bibinfo[2]{#2}
\providecommand\natexlab[1]{#1}
\providecommand\showeprint[2][]{arXiv:#2}

\bibitem[Alabdulwahhab(2018)]%
        {alabdulwahhab2018web}
\bibfield{author}{\bibinfo{person}{Faten~Adel Alabdulwahhab}.}
  \bibinfo{year}{2018}\natexlab{}.
\newblock \showarticletitle{Web 3.0: the decentralized web blockchain networks
  and protocol innovation}. In \bibinfo{booktitle}{\emph{2018 1st International
  Conference on Computer Applications \& Information Security (ICCAIS)}}. IEEE,
  \bibinfo{pages}{1--4}.
\newblock


\bibitem[Allombert et~al\mbox{.}(2019)]%
        {allombert2019introduction}
\bibfield{author}{\bibinfo{person}{Victor Allombert}, \bibinfo{person}{Mathias
  Bourgoin}, {and} \bibinfo{person}{Julien Tesson}.}
  \bibinfo{year}{2019}\natexlab{}.
\newblock \showarticletitle{Introduction to the tezos blockchain}. In
  \bibinfo{booktitle}{\emph{2019 International Conference on High Performance
  Computing \& Simulation (HPCS)}}. IEEE, \bibinfo{pages}{1--10}.
\newblock


\bibitem[Ba et~al\mbox{.}(2022)]%
        {ba2022fork}
\bibfield{author}{\bibinfo{person}{Cheick~Tidiane Ba}, \bibinfo{person}{Andrea
  Michienzi}, \bibinfo{person}{Barbara Guidi}, \bibinfo{person}{Matteo
  Zignani}, \bibinfo{person}{Laura Ricci}, {and} \bibinfo{person}{Sabrina
  Gaito}.} \bibinfo{year}{2022}\natexlab{}.
\newblock \showarticletitle{Fork-based user migration in blockchain online
  social media}. In \bibinfo{booktitle}{\emph{14th ACM Web Science Conference
  2022}}. \bibinfo{pages}{174--184}.
\newblock


\bibitem[Castro et~al\mbox{.}(1999)]%
        {castro1999practical}
\bibfield{author}{\bibinfo{person}{Miguel Castro}, \bibinfo{person}{Barbara
  Liskov}, {et~al\mbox{.}}} \bibinfo{year}{1999}\natexlab{}.
\newblock \showarticletitle{Practical byzantine fault tolerance}. In
  \bibinfo{booktitle}{\emph{OsDI}}, Vol.~\bibinfo{volume}{99}.
  \bibinfo{pages}{173--186}.
\newblock


\bibitem[Chen et~al\mbox{.}(2022)]%
        {chen2022digital}
\bibfield{author}{\bibinfo{person}{Chuan Chen}, \bibinfo{person}{Lei Zhang},
  \bibinfo{person}{Yihao Li}, \bibinfo{person}{Tianchi Liao},
  \bibinfo{person}{Siran Zhao}, \bibinfo{person}{Zibin Zheng},
  \bibinfo{person}{Huawei Huang}, {and} \bibinfo{person}{Jiajing Wu}.}
  \bibinfo{year}{2022}\natexlab{}.
\newblock \showarticletitle{When digital economy meets web 3.0: Applications
  and challenges}.
\newblock \bibinfo{journal}{\emph{IEEE Open Journal of the Computer Society}}
  (\bibinfo{year}{2022}).
\newblock


\bibitem[Croman et~al\mbox{.}(2016)]%
        {croman2016scaling}
\bibfield{author}{\bibinfo{person}{Kyle Croman} {et~al\mbox{.}}}
  \bibinfo{year}{2016}\natexlab{}.
\newblock \showarticletitle{On scaling decentralized blockchains}. In
  \bibinfo{booktitle}{\emph{International Conference on Financial Cryptography
  and Data Security}}. \bibinfo{pages}{106--125}.
\newblock


\bibitem[Franks and Mayer(1996)]%
        {franks1996hostile}
\bibfield{author}{\bibinfo{person}{Julian Franks} {and} \bibinfo{person}{Colin
  Mayer}.} \bibinfo{year}{1996}\natexlab{}.
\newblock \showarticletitle{Hostile takeovers and the correction of managerial
  failure}.
\newblock \bibinfo{journal}{\emph{Journal of financial economics}}
  \bibinfo{volume}{40}, \bibinfo{number}{1} (\bibinfo{year}{1996}),
  \bibinfo{pages}{163--181}.
\newblock


\bibitem[Gencer et~al\mbox{.}(2018)]%
        {gencer2018decentralization}
\bibfield{author}{\bibinfo{person}{Adem~Efe Gencer}, \bibinfo{person}{Soumya
  Basu}, \bibinfo{person}{Ittay Eyal}, \bibinfo{person}{Robbert Van~Renesse},
  {and} \bibinfo{person}{Emin~G{\"u}n Sirer}.} \bibinfo{year}{2018}\natexlab{}.
\newblock \showarticletitle{Decentralization in bitcoin and ethereum networks}.
  In \bibinfo{booktitle}{\emph{International Conference on Financial
  Cryptography and Data Security}}. Springer, \bibinfo{pages}{439--457}.
\newblock


\bibitem[Gervais et~al\mbox{.}(2014)]%
        {gervais2014bitcoin}
\bibfield{author}{\bibinfo{person}{Arthur Gervais}, \bibinfo{person}{Ghassan~O
  Karame}, \bibinfo{person}{Vedran Capkun}, {and} \bibinfo{person}{Srdjan
  Capkun}.} \bibinfo{year}{2014}\natexlab{}.
\newblock \showarticletitle{Is bitcoin a decentralized currency?}
\newblock \bibinfo{journal}{\emph{IEEE security \& privacy}}
  \bibinfo{volume}{12}, \bibinfo{number}{3} (\bibinfo{year}{2014}),
  \bibinfo{pages}{54--60}.
\newblock


\bibitem[Guidi(2020)]%
        {guidi2020blockchain}
\bibfield{author}{\bibinfo{person}{Barbara Guidi}.}
  \bibinfo{year}{2020}\natexlab{}.
\newblock \showarticletitle{When blockchain meets online social networks}.
\newblock \bibinfo{journal}{\emph{Pervasive and Mobile Computing}}
  \bibinfo{volume}{62} (\bibinfo{year}{2020}), \bibinfo{pages}{101131}.
\newblock


\bibitem[Guidi et~al\mbox{.}(2020a)]%
        {guidi2020graph}
\bibfield{author}{\bibinfo{person}{Barbara Guidi}, \bibinfo{person}{Andrea
  Michienzi}, {and} \bibinfo{person}{Laura Ricci}.}
  \bibinfo{year}{2020}\natexlab{a}.
\newblock \showarticletitle{A graph-based socioeconomic analysis of steemit}.
\newblock \bibinfo{journal}{\emph{IEEE Transactions on Computational Social
  Systems}} \bibinfo{volume}{8}, \bibinfo{number}{2} (\bibinfo{year}{2020}),
  \bibinfo{pages}{365--376}.
\newblock


\bibitem[Guidi et~al\mbox{.}(2020b)]%
        {guidi2020steem}
\bibfield{author}{\bibinfo{person}{Barbara Guidi}, \bibinfo{person}{Andrea
  Michienzi}, {and} \bibinfo{person}{Laura Ricci}.}
  \bibinfo{year}{2020}\natexlab{b}.
\newblock \showarticletitle{Steem blockchain: Mining the inner structure of the
  graph}.
\newblock \bibinfo{journal}{\emph{IEEE Access}}  \bibinfo{volume}{8}
  (\bibinfo{year}{2020}), \bibinfo{pages}{210251--210266}.
\newblock


\bibitem[Guidi et~al\mbox{.}(2021)]%
        {guidi2021analysis}
\bibfield{author}{\bibinfo{person}{Barbara Guidi}, \bibinfo{person}{Andrea
  Michienzi}, {and} \bibinfo{person}{Laura Ricci}.}
  \bibinfo{year}{2021}\natexlab{}.
\newblock \showarticletitle{Analysis of witnesses in the steem blockchain}.
\newblock \bibinfo{journal}{\emph{Mobile Networks and Applications}}
  (\bibinfo{year}{2021}), \bibinfo{pages}{1--12}.
\newblock


\bibitem[Guidi et~al\mbox{.}(2022)]%
        {guidi2022assessment}
\bibfield{author}{\bibinfo{person}{Barbara Guidi}, \bibinfo{person}{Andrea
  Michienzi}, {and} \bibinfo{person}{Laura Ricci}.}
  \bibinfo{year}{2022}\natexlab{}.
\newblock \showarticletitle{Assessment of Wealth Distribution in Blockchain
  Online Social Media}.
\newblock \bibinfo{journal}{\emph{IEEE Transactions on Computational Social
  Systems}} (\bibinfo{year}{2022}).
\newblock


\bibitem[{Hive Block Explorer [Internet]}(2023)]%
        {Hive}
\bibfield{author}{\bibinfo{person}{{Hive Block Explorer [Internet]}}.}
  \bibinfo{year}{2023}\natexlab{}.
\newblock \bibinfo{howpublished}{\url{https://www.hiveblockexplorer.com/}}.
\newblock
\newblock
\shownote{Accessed Dec. 2023}.


\bibitem[{Interactive Hive API [Internet]}(2023)]%
        {HiveAPI}
\bibfield{author}{\bibinfo{person}{{Interactive Hive API [Internet]}}.}
  \bibinfo{year}{2023}\natexlab{}.
\newblock \bibinfo{howpublished}{\url{https://developers.hive.io/}}.
\newblock
\newblock
\shownote{Accessed Dec. 2023}.


\bibitem[{Interactive Steem API [Internet]}(2023)]%
        {SteemAPI}
\bibfield{author}{\bibinfo{person}{{Interactive Steem API [Internet]}}.}
  \bibinfo{year}{2023}\natexlab{}.
\newblock \bibinfo{howpublished}{\url{https://developers.steem.io/}}.
\newblock
\newblock
\shownote{Accessed Dec. 2023}.


\bibitem[Jeong(2020)]%
        {jeong2020centralized}
\bibfield{author}{\bibinfo{person}{Seungwon~Eugene Jeong}.}
  \bibinfo{year}{2020}\natexlab{}.
\newblock \showarticletitle{Centralized decentralization: Does voting matter?
  simple economics of the dpos blockchain governance}.
\newblock \bibinfo{journal}{\emph{Simple Economics of the DPoS Blockchain
  Governance (April 21, 2020)}} (\bibinfo{year}{2020}).
\newblock


\bibitem[Kiayias and Lazos(2022)]%
        {kiayias2022sok}
\bibfield{author}{\bibinfo{person}{Aggelos Kiayias} {and}
  \bibinfo{person}{Philip Lazos}.} \bibinfo{year}{2022}\natexlab{}.
\newblock \showarticletitle{SoK: blockchain governance}.
\newblock \bibinfo{journal}{\emph{arXiv preprint arXiv:2201.07188}}
  (\bibinfo{year}{2022}).
\newblock


\bibitem[King and Nadal(2012)]%
        {king2012ppcoin}
\bibfield{author}{\bibinfo{person}{Sunny King} {and} \bibinfo{person}{Scott
  Nadal}.} \bibinfo{year}{2012}\natexlab{}.
\newblock \showarticletitle{Ppcoin: Peer-to-peer crypto-currency with
  proof-of-stake}.
\newblock \bibinfo{journal}{\emph{self-published paper, August}}
  \bibinfo{volume}{19}, \bibinfo{number}{1} (\bibinfo{year}{2012}).
\newblock


\bibitem[Kwon et~al\mbox{.}(2019)]%
        {kwon2019impossibility}
\bibfield{author}{\bibinfo{person}{Yujin Kwon}, \bibinfo{person}{Jian Liu},
  \bibinfo{person}{Minjeong Kim}, \bibinfo{person}{Dawn Song}, {and}
  \bibinfo{person}{Yongdae Kim}.} \bibinfo{year}{2019}\natexlab{}.
\newblock \showarticletitle{Impossibility of full decentralization in
  permissionless blockchains}. In \bibinfo{booktitle}{\emph{Proceedings of the
  1st ACM Conference on Advances in Financial Technologies}}.
  \bibinfo{pages}{110--123}.
\newblock


\bibitem[Larimer(2014)]%
        {larimer2014delegated}
\bibfield{author}{\bibinfo{person}{Daniel Larimer}.}
  \bibinfo{year}{2014}\natexlab{}.
\newblock \showarticletitle{Delegated proof-of-stake (dpos)}.
\newblock \bibinfo{journal}{\emph{Bitshare whitepaper}} (\bibinfo{year}{2014}).
\newblock


\bibitem[Leiponen et~al\mbox{.}(2022)]%
        {leiponen2022dapp}
\bibfield{author}{\bibinfo{person}{Aija Leiponen},
  \bibinfo{person}{Llewellyn~DW Thomas}, {and} \bibinfo{person}{Qian Wang}.}
  \bibinfo{year}{2022}\natexlab{}.
\newblock \showarticletitle{The dApp economy: a new platform for distributed
  innovation?}
\newblock \bibinfo{journal}{\emph{Innovation}} \bibinfo{volume}{24},
  \bibinfo{number}{1} (\bibinfo{year}{2022}), \bibinfo{pages}{125--143}.
\newblock


\bibitem[Li and Palanisamy(2019)]%
        {li2019incentivized}
\bibfield{author}{\bibinfo{person}{Chao Li} {and} \bibinfo{person}{Balaji
  Palanisamy}.} \bibinfo{year}{2019}\natexlab{}.
\newblock \showarticletitle{Incentivized blockchain-based social media
  platforms: A case study of steemit}. In \bibinfo{booktitle}{\emph{Proceedings
  of the 10th ACM conference on web science}}. \bibinfo{pages}{145--154}.
\newblock


\bibitem[Li and Palanisamy(2020)]%
        {li2020comparison}
\bibfield{author}{\bibinfo{person}{Chao Li} {and} \bibinfo{person}{Balaji
  Palanisamy}.} \bibinfo{year}{2020}\natexlab{}.
\newblock \showarticletitle{Comparison of decentralization in dpos and pow
  blockchains}. In \bibinfo{booktitle}{\emph{International Conference on
  Blockchain}}. Springer, \bibinfo{pages}{18--32}.
\newblock


\bibitem[Li et~al\mbox{.}(2023a)]%
        {li2023cross}
\bibfield{author}{\bibinfo{person}{Chao Li}, \bibinfo{person}{Balaji
  Palanisamy}, \bibinfo{person}{Runhua Xu}, {and} \bibinfo{person}{Li Duan}.}
  \bibinfo{year}{2023}\natexlab{a}.
\newblock \showarticletitle{Cross-Consensus Measurement of Individual-level
  Decentralization in Blockchains}. In \bibinfo{booktitle}{\emph{2023 IEEE 9th
  Intl Conference on Big Data Security on Cloud (BigDataSecurity), IEEE Intl
  Conference on High Performance and Smart Computing,(HPSC) and IEEE Intl
  Conference on Intelligent Data and Security (IDS)}}. IEEE,
  \bibinfo{pages}{45--50}.
\newblock


\bibitem[Li et~al\mbox{.}(2023b)]%
        {li2023hard}
\bibfield{author}{\bibinfo{person}{Chao Li}, \bibinfo{person}{Balaji
  Palanisamy}, \bibinfo{person}{Runhua Xu}, \bibinfo{person}{Li Duan},
  \bibinfo{person}{Jiqiang Liu}, {and} \bibinfo{person}{Wei Wang}.}
  \bibinfo{year}{2023}\natexlab{b}.
\newblock \showarticletitle{How Hard is Takeover in DPoS Blockchains?
  Understanding the Security of Coin-based Voting Governance}. In
  \bibinfo{booktitle}{\emph{Proceedings of the 2023 ACM SIGSAC Conference on
  Computer and Communications Security}}. \bibinfo{pages}{150--164}.
\newblock


\bibitem[Li et~al\mbox{.}(2021)]%
        {li2021steemops}
\bibfield{author}{\bibinfo{person}{Chao Li}, \bibinfo{person}{Balaji
  Palanisamy}, \bibinfo{person}{Runhua Xu}, \bibinfo{person}{Jinlai Xu}, {and}
  \bibinfo{person}{Jingzhe Wang}.} \bibinfo{year}{2021}\natexlab{}.
\newblock \showarticletitle{Steemops: Extracting and analyzing key operations
  in steemit blockchain-based social media platform}. In
  \bibinfo{booktitle}{\emph{Proceedings of the Eleventh ACM Conference on Data
  and Application Security and Privacy}}. \bibinfo{pages}{113--118}.
\newblock


\bibitem[Li et~al\mbox{.}(2023c)]%
        {li2023characterizing}
\bibfield{author}{\bibinfo{person}{Chao Li}, \bibinfo{person}{Runhua Xu}, {and}
  \bibinfo{person}{Li Duan}.} \bibinfo{year}{2023}\natexlab{c}.
\newblock \showarticletitle{Characterizing Coin-Based Voting Governance in DPoS
  Blockchains}. In \bibinfo{booktitle}{\emph{Proceedings of the International
  AAAI Conference on Web and Social Media}}, Vol.~\bibinfo{volume}{17}.
  \bibinfo{pages}{1148--1152}.
\newblock


\bibitem[Li et~al\mbox{.}(2023d)]%
        {li2023liquid}
\bibfield{author}{\bibinfo{person}{Chao Li}, \bibinfo{person}{Runhua Xu}, {and}
  \bibinfo{person}{Li Duan}.} \bibinfo{year}{2023}\natexlab{d}.
\newblock \showarticletitle{Liquid democracy in DPoS blockchains}. In
  \bibinfo{booktitle}{\emph{Proceedings of the 5th ACM International Symposium
  on Blockchain and Secure Critical Infrastructure}}. \bibinfo{pages}{25--33}.
\newblock


\bibitem[Lin et~al\mbox{.}(2021)]%
        {lin2021measuring}
\bibfield{author}{\bibinfo{person}{Qinwei Lin}, \bibinfo{person}{Chao Li},
  \bibinfo{person}{Xifeng Zhao}, {and} \bibinfo{person}{Xianhai Chen}.}
  \bibinfo{year}{2021}\natexlab{}.
\newblock \showarticletitle{Measuring decentralization in bitcoin and ethereum
  using multiple metrics and granularities}. In \bibinfo{booktitle}{\emph{2021
  IEEE 37th International Conference on Data Engineering Workshops (ICDEW)}}.
  IEEE, \bibinfo{pages}{80--87}.
\newblock


\bibitem[Liu et~al\mbox{.}(2023)]%
        {liu2023demonitor}
\bibfield{author}{\bibinfo{person}{Jingyu Liu}, \bibinfo{person}{Lu Liu},
  \bibinfo{person}{Zijing Li}, {and} \bibinfo{person}{Chao Li}.}
  \bibinfo{year}{2023}\natexlab{}.
\newblock \showarticletitle{DeMonitor: Monitoring Decentralization in
  Blockchains using BigQuery}. In \bibinfo{booktitle}{\emph{2023 IEEE
  International Conference on Blockchain and Cryptocurrency (ICBC)}}. IEEE,
  \bibinfo{pages}{1--2}.
\newblock


\bibitem[Liu et~al\mbox{.}(2022)]%
        {liu2022decentralization}
\bibfield{author}{\bibinfo{person}{Jieli Liu}, \bibinfo{person}{Weilin Zheng},
  \bibinfo{person}{Dingyuan Lu}, \bibinfo{person}{Jiajing Wu}, {and}
  \bibinfo{person}{Zibin Zheng}.} \bibinfo{year}{2022}\natexlab{}.
\newblock \showarticletitle{From Decentralization to Oligopoly: A Data-Driven
  Analysis of Decentralization Evolution and Voting Behaviors on EOSIO}.
\newblock \bibinfo{journal}{\emph{IEEE Transactions on Computational Social
  Systems}} (\bibinfo{year}{2022}).
\newblock


\bibitem[Miller et~al\mbox{.}(2015)]%
        {miller2015discovering}
\bibfield{author}{\bibinfo{person}{Andrew Miller}, \bibinfo{person}{James
  Litton}, \bibinfo{person}{Andrew Pachulski}, \bibinfo{person}{Neal Gupta},
  \bibinfo{person}{Dave Levin}, \bibinfo{person}{Neil Spring}, {and}
  \bibinfo{person}{Bobby Bhattacharjee}.} \bibinfo{year}{2015}\natexlab{}.
\newblock \showarticletitle{Discovering bitcoin’s public topology and
  influential nodes}.
\newblock \bibinfo{journal}{\emph{et al}} (\bibinfo{year}{2015}).
\newblock


\bibitem[Nakamoto(2008)]%
        {nakamoto2008bitcoin}
\bibfield{author}{\bibinfo{person}{Satoshi Nakamoto}.}
  \bibinfo{year}{2008}\natexlab{}.
\newblock \showarticletitle{Bitcoin: A peer-to-peer electronic cash system}.
\newblock  (\bibinfo{year}{2008}).
\newblock


\bibitem[Ragnedda and Destefanis(2019)]%
        {ragnedda2019blockchain}
\bibfield{author}{\bibinfo{person}{Massimo Ragnedda} {and}
  \bibinfo{person}{Giuseppe Destefanis}.} \bibinfo{year}{2019}\natexlab{}.
\newblock \bibinfo{booktitle}{\emph{Blockchain and Web 3.0}}.
\newblock \bibinfo{publisher}{London: Routledge, Taylor and Francis Group}.
\newblock


\bibitem[Shen et~al\mbox{.}(2024)]%
        {shen2024artificial}
\bibfield{author}{\bibinfo{person}{Meng Shen}, \bibinfo{person}{Zhehui Tan},
  \bibinfo{person}{Dusit Niyato}, \bibinfo{person}{Yuzhi Liu},
  \bibinfo{person}{Jiawen Kang}, \bibinfo{person}{Zehui Xiong},
  \bibinfo{person}{Liehuang Zhu}, \bibinfo{person}{Wei Wang}, {and}
  \bibinfo{person}{Xuemin Shen}.} \bibinfo{year}{2024}\natexlab{}.
\newblock \showarticletitle{Artificial Intelligence for Web 3.0: A
  Comprehensive Survey}.
\newblock \bibinfo{journal}{\emph{Comput. Surveys}} \bibinfo{volume}{56},
  \bibinfo{number}{10} (\bibinfo{year}{2024}), \bibinfo{pages}{1--39}.
\newblock


\bibitem[Shi et~al\mbox{.}(2023)]%
        {shi2023ress}
\bibfield{author}{\bibinfo{person}{Dongxian Shi}, \bibinfo{person}{Xiaoqing
  Wang}, \bibinfo{person}{Ming Xu}, \bibinfo{person}{Liang Kou}, {and}
  \bibinfo{person}{Hongbing Cheng}.} \bibinfo{year}{2023}\natexlab{}.
\newblock \showarticletitle{Ress: A reliable and effcient storage scheme for
  bitcoin blockchain based on raptor code}.
\newblock \bibinfo{journal}{\emph{Chinese Journal of Electronics}}
  \bibinfo{volume}{32}, \bibinfo{number}{3} (\bibinfo{year}{2023}),
  \bibinfo{pages}{577--586}.
\newblock


\bibitem[Shivdasani(1993)]%
        {shivdasani1993board}
\bibfield{author}{\bibinfo{person}{Anil Shivdasani}.}
  \bibinfo{year}{1993}\natexlab{}.
\newblock \showarticletitle{Board composition, ownership structure, and hostile
  takeovers}.
\newblock \bibinfo{journal}{\emph{Journal of accounting and economics}}
  \bibinfo{volume}{16}, \bibinfo{number}{1-3} (\bibinfo{year}{1993}),
  \bibinfo{pages}{167--198}.
\newblock


\bibitem[Su and Jiang(2023)]%
        {su2023hybrid}
\bibfield{author}{\bibinfo{person}{Jian Su} {and} \bibinfo{person}{Mengnan
  Jiang}.} \bibinfo{year}{2023}\natexlab{}.
\newblock \showarticletitle{A hybrid entropy and blockchain approach for
  network security defense in SDN-based IIoT}.
\newblock \bibinfo{journal}{\emph{Chinese Journal of Electronics}}
  \bibinfo{volume}{32}, \bibinfo{number}{3} (\bibinfo{year}{2023}),
  \bibinfo{pages}{531--541}.
\newblock


\bibitem[{TRON white paper [Internet]}(2023)]%
        {tron}
\bibfield{author}{\bibinfo{person}{{TRON white paper [Internet]}}.}
  \bibinfo{year}{2023}\natexlab{}.
\newblock
  \bibinfo{howpublished}{\url{https://tron.network/static/doc/white_paper_v_2_0.pdf/}}.
\newblock
\newblock
\shownote{Accessed Dec. 2023}.


\bibitem[Wang et~al\mbox{.}(2020)]%
        {wang2020measurement}
\bibfield{author}{\bibinfo{person}{Canhui Wang}, \bibinfo{person}{Xiaowen Chu},
  {and} \bibinfo{person}{Yang Qin}.} \bibinfo{year}{2020}\natexlab{}.
\newblock \showarticletitle{Measurement and analysis of the bitcoin networks: A
  view from mining pools}. In \bibinfo{booktitle}{\emph{2020 6th International
  Conference on Big Data Computing and Communications (BIGCOM)}}. IEEE,
  \bibinfo{pages}{180--188}.
\newblock


\bibitem[Wang et~al\mbox{.}(2022)]%
        {wang2022exploring}
\bibfield{author}{\bibinfo{person}{Qin Wang}, \bibinfo{person}{Rujia Li},
  \bibinfo{person}{Qi Wang}, \bibinfo{person}{Shiping Chen}, {and}
  \bibinfo{person}{Yang Xiang}.} \bibinfo{year}{2022}\natexlab{}.
\newblock \showarticletitle{Exploring unfairness on proof of authority: Order
  manipulation attacks and remedies}. In \bibinfo{booktitle}{\emph{Proceedings
  of the 2022 ACM on Asia Conference on Computer and Communications Security}}.
  \bibinfo{pages}{123--137}.
\newblock


\bibitem[Wood(2014)]%
        {wood2014ethereum}
\bibfield{author}{\bibinfo{person}{Gavin Wood}.}
  \bibinfo{year}{2014}\natexlab{}.
\newblock \showarticletitle{Ethereum: A secure decentralised generalised
  transaction ledger}.
\newblock \bibinfo{journal}{\emph{Ethereum project yellow paper}}
  \bibinfo{volume}{151} (\bibinfo{year}{2014}), \bibinfo{pages}{1--32}.
\newblock


\bibitem[Wood(2016)]%
        {wood2016polkadot}
\bibfield{author}{\bibinfo{person}{Gavin Wood}.}
  \bibinfo{year}{2016}\natexlab{}.
\newblock \showarticletitle{Polkadot: Vision for a heterogeneous multi-chain
  framework}.
\newblock \bibinfo{journal}{\emph{White paper}} \bibinfo{volume}{21},
  \bibinfo{number}{2327} (\bibinfo{year}{2016}), \bibinfo{pages}{4662}.
\newblock


\bibitem[Wu et~al\mbox{.}(2023)]%
        {wu2023financial}
\bibfield{author}{\bibinfo{person}{Jiajing Wu}, \bibinfo{person}{Kaixin Lin},
  \bibinfo{person}{Dan Lin}, \bibinfo{person}{Ziye Zheng},
  \bibinfo{person}{Huawei Huang}, {and} \bibinfo{person}{Zibin Zheng}.}
  \bibinfo{year}{2023}\natexlab{}.
\newblock \showarticletitle{Financial Crimes in Web3-empowered Metaverse:
  Taxonomy, Countermeasures, and Opportunities}.
\newblock \bibinfo{journal}{\emph{IEEE Open Journal of the Computer Society}}
  \bibinfo{volume}{4} (\bibinfo{year}{2023}), \bibinfo{pages}{37--49}.
\newblock


\bibitem[Wu et~al\mbox{.}(2019)]%
        {wu2019information}
\bibfield{author}{\bibinfo{person}{Keke Wu}, \bibinfo{person}{Bo Peng},
  \bibinfo{person}{Hua Xie}, {and} \bibinfo{person}{Zhen Huang}.}
  \bibinfo{year}{2019}\natexlab{}.
\newblock \showarticletitle{An information entropy method to quantify the
  degrees of decentralization for blockchain systems}. In
  \bibinfo{booktitle}{\emph{2019 IEEE 9th International Conference on
  Electronics Information and Emergency Communication (ICEIEC)}}. IEEE,
  \bibinfo{pages}{1--6}.
\newblock


\bibitem[Yang et~al\mbox{.}(2023)]%
        {yang2023zero}
\bibfield{author}{\bibinfo{person}{Kunwei Yang}, \bibinfo{person}{Bo Yang},
  \bibinfo{person}{Tao Wang}, {and} \bibinfo{person}{Yanwei Zhou}.}
  \bibinfo{year}{2023}\natexlab{}.
\newblock \showarticletitle{Zero-Cerd: A Self-Blindable Anonymous
  Authentication System Based on Blockchain}.
\newblock \bibinfo{journal}{\emph{Chinese Journal of Electronics}}
  \bibinfo{volume}{32}, \bibinfo{number}{3} (\bibinfo{year}{2023}),
  \bibinfo{pages}{587--596}.
\newblock


\bibitem[Zeng et~al\mbox{.}(2021)]%
        {zeng2021characterizing}
\bibfield{author}{\bibinfo{person}{Liyi Zeng}, \bibinfo{person}{Yang Chen},
  \bibinfo{person}{Shuo Chen}, \bibinfo{person}{Xian Zhang},
  \bibinfo{person}{Zhongxin Guo}, \bibinfo{person}{Wei Xu}, {and}
  \bibinfo{person}{Thomas Moscibroda}.} \bibinfo{year}{2021}\natexlab{}.
\newblock \showarticletitle{Characterizing Ethereum’s Mining Power
  Decentralization at a Deeper Level}. In \bibinfo{booktitle}{\emph{IEEE
  INFOCOM 2021-IEEE Conference on Computer Communications}}. IEEE,
  \bibinfo{pages}{1--10}.
\newblock


\bibitem[Zheng et~al\mbox{.}(2023a)]%
        {zheng2023blockchain}
\bibfield{author}{\bibinfo{person}{Peilin Zheng}, \bibinfo{person}{Zigui
  Jiang}, \bibinfo{person}{Jiajing Wu}, {and} \bibinfo{person}{Zibin Zheng}.}
  \bibinfo{year}{2023}\natexlab{a}.
\newblock \showarticletitle{Blockchain-based decentralized application: A
  survey}.
\newblock \bibinfo{journal}{\emph{IEEE Open Journal of the Computer Society}}
  (\bibinfo{year}{2023}).
\newblock


\bibitem[Zheng et~al\mbox{.}(2023b)]%
        {zheng2023bshunter}
\bibfield{author}{\bibinfo{person}{Peilin Zheng}, \bibinfo{person}{Xiapu Luo},
  {and} \bibinfo{person}{Zibin Zheng}.} \bibinfo{year}{2023}\natexlab{b}.
\newblock \showarticletitle{BSHUNTER: Detecting and Tracing Defects of Bitcoin
  Scripts}. In \bibinfo{booktitle}{\emph{2023 IEEE/ACM 45th International
  Conference on Software Engineering (ICSE)}}. IEEE, \bibinfo{pages}{307--318}.
\newblock


\bibitem[Zheng et~al\mbox{.}(2020)]%
        {zheng2020xblock}
\bibfield{author}{\bibinfo{person}{Peilin Zheng}, \bibinfo{person}{Zibin
  Zheng}, \bibinfo{person}{Jiajing Wu}, {and} \bibinfo{person}{Hong-Ning Dai}.}
  \bibinfo{year}{2020}\natexlab{}.
\newblock \showarticletitle{Xblock-eth: Extracting and exploring blockchain
  data from ethereum}.
\newblock \bibinfo{journal}{\emph{IEEE Open Journal of the Computer Society}}
  \bibinfo{volume}{1} (\bibinfo{year}{2020}), \bibinfo{pages}{95--106}.
\newblock


\end{thebibliography}

\end{document}